\documentclass[12pt,twoside]{article}
\usepackage[mathscr]{eucal}
\usepackage{amsmath,amsfonts,amssymb,amsthm}
\usepackage{hyperref}
\usepackage{times}
\usepackage[nosort]{cite}
\def\d_Vphi{\text{d}_V\hspace{-0.06em}\phi}
\def\d_Vphibar{\text{d}_V\hspace{-0.06em}\bar\phi}
\def\d_Vxi{\text{d}_V\hspace{-0.06em}\xi}

\def\sA{\mathbf A}
\def\sD{\mathbf D}

\def\be{\begin{eqnarray}}
\def\ee{\end{eqnarray}}
\def\beann{\begin{eqnarray*}}
\def\eeann{\end{eqnarray*}}
\def\beq{\begin{equation}}
\def\eeq{\end{equation}}
\def\ba{\begin{array}}
\def\ea{\end{array}}
\def\ben{\begin{enumerate}}
\def\een{\end{enumerate}}
\def\bea{\begin{eqnarray}}
\def\eea{\end{eqnarray}}

\def\5{\bar }
\def\6{\partial }
\def\7{\hat }
\def\4{\tilde }
\advance\textheight\topskip
\renewcommand{\tilde}{\widetilde}
\renewcommand{\hat}{\widehat}
\theoremstyle{plain}
\newtheorem{prop}{Proposition}
\newtheorem{lemma}[prop]{Lemma}

\renewcommand{\simeq}{\cong}

\newcommand{\bref}[1]{\textbf{\ref{#1}}}

\newcommand{\Ker}{\mathop{\mathrm{Ker}}}
\newcommand{\im}{\mathop{\mathrm{Im}}}

\newcommand{\p}[1]{|#1|}
\newcommand{\gh}[1]{\mathrm{gh}(#1)}

\newcommand{\assalgebra}{\mathscr}    %ass[ociative] algebra(s)
\newcommand{\algA}{\assalgA}
\newcommand{\assalgA}{\assalgebra{A}}

\newcommand{\dd}{\partial}
\renewcommand{\d}{\partial}

\renewcommand{\geq}{\,{\geqslant}\,}
\renewcommand{\leq}{\,{\leqslant}\,}

\newcommand{\inner}[2]{\langle #1{,}\,#2\rangle}
\newcommand{\binner}[2]{%
  {\langle}\kern-4.15pt{\langle}#1{,}\,#2{\rangle}\kern-4.15pt{\rangle}}
\newcommand{\commut}[2]{[#1{,}\,#2]}

\newcommand{\ab}[2]{\big(#1,#2\big)}

\newcommand{\half}{\mathchoice{%
    \ffrac{1}{2}}{\frac{1}{2}}{\frac{1}{2}}{\frac{1}{2}}}

\newcommand{\ffrac}[2]{\raisebox{.5pt}%
  {\footnotesize$\displaystyle\frac{#1}{#2}$}\kern1pt}

\newcommand{\brst}{\mathsf{\Omega}}
\newcommand{\Qbrst}{\mathsf{\Omega}}
\newcommand{\Qred}{{Q}}

\newcommand{\dl}[1]{\mathchoice{\ffrac{\dd}{\dd #1}}{\frac{\dd}{\dd
      #1}}{\ffrac{\dd}{\dd #1}}{\ffrac{\dd}{\dd #1}}}

\newcommand{\ddl}[2]{\ffrac{\dd #1}{\dd #2}}

\newcommand{\vddr}[2]{\ffrac{\delta^R #1}{\delta #2}}

\newcommand{\vddll}[2]{{\ffrac{\delta^L #1}{\delta #2}}}

\newcommand{\os}{{a}}%\dagger}}
\newcommand{\Os}{{A}}%\dagger}}
\newcommand{\CG}{{\Xi}}
\newcommand{\fR}{\mathbb{R}}

\def\cD{\mathcal{D}}

\def\cG{\mathcal{G}}
\def\cH{\mathcal{H}}

\def\cK{\mathcal{K}}
\def\cL{\mathcal{L}}

\def\cP{\mathcal{P}}

\def\cS{\mathcal{S}}
\def\cT{\mathcal{T}}

\numberwithin{equation}{section} \makeatletter
\@addtoreset{equation}{section}

\begin{document}

\def\mytitle{Notes on conformal invariance of gauge fields}

\pagestyle{myheadings} \markboth{\textsc{\small Barnich, Bekaert, Grigoriev}}{%
  \textsc{\small Conformal invariance of free higher spin gauge fields}}
\addtolength{\headsep}{4pt}

\begin{centering}

  \vspace{1cm}

  \textbf{\Large{\mytitle}}

%\vspace{1cm}

%{\huge Notes}

  \vspace{1.5cm}

  {\large Glenn Barnich} 

\vspace{.5cm}

\begin{minipage}{.9\textwidth}\small \it \begin{center}
    Physique Th\'eorique et Math\'ematique\\ Universit\'e Libre de
    Bruxelles and International Solvay Institutes \\ Campus
    Plaine C.P. 231, B-1050 Bruxelles, Belgium \end{center}
\end{minipage}

\vspace{.5cm}

 {\large Xavier Bekaert} 

\vspace{.5cm}

\begin{minipage}{.9\textwidth}\small \it \begin{center}
    Laboratoire de Math\'ematiques et Physique Th\'eorique\\
{Unit\'e Mixte de Recherche $7350$ du CNRS\\
F\'ed\'eration de Recherche $2964$ Denis Poisson}\\
Universit\'e Fran\c cois Rabelais\\
Parc de Grandmont,
37200 Tours, France 
  \end{center}
\end{minipage}

\vspace{.5cm}

 {\large  Maxim Grigoriev}

\vspace{.5cm}

\begin{minipage}{.9\textwidth}\small \it \begin{center}
    I.E. Tamm Department of Theoretical Physics\\
    P.N. Lebedev Physical Institute\\
    Leninsky prospect 53, 119991 Moscow, Russia\end{center}
\end{minipage}

\end{centering}

\vspace{1cm}

\begin{center}
  \begin{minipage}{.9\textwidth}
    \textsc{Abstract}. In Lagrangian gauge systems, the vector space
    of global reducibility parameters forms a module under the Lie
    algebra of symmetries of the action. Since the classification of
    global reducibility parameters is generically easier than the
    classification of symmetries of the action, this fact can be used
    to constrain the latter when knowing the former. We apply this
    strategy and its generalization for the non-Lagrangian setting to
    the problem of conformal symmetry of various free higher spin
    gauge fields. This scheme allows one to show that, in terms of
    potentials, massless higher spin gauge fields in Minkowski space
    and partially-massless fields in (A)dS space are not conformal for
    spin strictly greater than one, while in terms of curvatures,
    maximal-depth partially-massless fields in four dimensions are
    also not conformal, unlike the closely related, but less
    constrained, maximal-depth Fradkin--Tseytlin fields.
  \end{minipage}
\end{center}

\thispagestyle{empty}
\newpage

\begin{small}
{\addtolength{\parskip}{-1.6pt}
 \tableofcontents}
\end{small}
%\newpage

\section{Generalities}
\label{sec:generalities}

\subsection{Plan of the paper}

In the first section, we briefly review relevant aspects of symmetries
in the context of gauge systems: variational versus equations of
motion symmetries, field-theoretic formulation of conformal symmetry,
curvature versus potential formulations, the BRST-BV implementation,
and the relation to the unfolded approach. We demonstrate that for a
gauge system invariant under a global symmetry algebra, the space of
global reducibility parameters, and more generally, certain BRST
cohomology groups, are necessarily a module thereof. This gives a
powerful criterion to analyze whether a given gauge system admits a
given global symmetry algebra.

In section~\bref{sec:conf-invar-fronsd} we apply this criterion to
generic gauge fields in Minkowski space. More precisely, we address
the question which general mixed-symmetry bosonic gauge fields on
Minkowski space admit an extension from Poincar\'e to conformal
symmetry. We also illustrate the difference between variational and
equations of motion symmetries using the simplest example of a
massless scalar.

Section~\bref{sec:AdSfields} is devoted to identifying those gauge
fields on anti-de Sitter (AdS) space whose AdS symmetry extends to
conformal symmetry. We pay particular attention to the special case of
maximal-depth partially-massless (PM) fields in $AdS_4$ because these
fields have attracted some attention in the literature and can easily
be confused with their conformal cousins belonging to the family of
(generalized) Fradkin--Tseytlin fields, which we also discuss. We show
that these fields are never conformal for $s>1$ neither as gauge
fields nor at the level of gauge invariant curvatures. As an
illustration the case of $s=2$ is considered in detail.

\subsection{Classification of symmetries}
\label{sec:class-symm}

Algebraic approaches to classifying symmetries of systems of partial
differential equations in the context of jet-bundles and the
variational bicomplex are by now very well-developed, see
e.g.~\cite{Vinogradov:1977,Vinogradov:1978,%
  Vinogradov:1984,vinogradov:2001} and also
\cite{Anderson1991,Dickey:1991xa,Olver:1993,Andersonbook,%
  Bluman2002,Bluman2009} for reviews. In particular for Lagrangian
systems, symmetries of the action, also called variational symmetries,
are a subalgebra of the symmetries of the equations of motion. In
applications to fundamental systems, they are privileged since
Noether's theorem provides one with a clear procedure on how to
implement them in the quantum theory.

The case of Lagrangian gauge systems and of degenerate partial
differential equations is less studied in the mathematical literature,
mainly because gauge invariance violates technical assumptions needed
to apply some of the systematic techniques (see however
\cite{Krasilshchik1998} and references therein).

For example, for massless higher-spin fields in four-dimensional flat
spacetime, symmetries and conservation laws of the equations of
motions have been classified in terms of curvatures
\cite{Anco:2002xn,Pohjanpelto:2008st} (see also \cite{Bekaert:2009fg} for
considerations in higher dimensions), generalizing the result for a
massless scalar field \cite{Eastwood:2002su}. With quantization in
mind (see e.g.~\cite{Fierz:1939ix} for an early discussion), suitable
potentials and auxiliary fields are introduced in order to make the
system Lagrangian, at the expense of introducing gauge symmetries in
the massless case \cite{Fronsdal:1978rb,Fang:1978wz}. A classification
of variational symmetries, and thus also of conservation laws, in such
formulations, would be very useful. In particular, one needs to
consider suitable equivalence classes of symmetries modulo gauge ones.

\subsection{Conformal symmetry}
\label{sec:conf-subalg-glob}

Short of a complete classification of symmetries, a standard question
is whether a given system admits certain subalgebras of
symmetries. Typically, in the situation that we consider below, the
relevant systems are by construction invariant under a certain
subalgebra of symmetries and one would like to know whether they admit
an extension to a bigger algebra of symmetries containing the starting
point algebra as a subalgebra. For a variety of field-theoretical
realizations of the Poincar\'e or the (anti-) de Sitter algebra for
instance, the role of the bigger algebra is played by the conformal
algebra.

This question has been thoroughly studied in two related -- but in
general not entirely equivalent -- approaches. The first one is purely
representation-theoretical and studies which (A)dS or Poincar{\'e}
irreps (usually unitary ones) can be lifted to irreps of the conformal
group~\cite{Gross1964,Bracken:1982ny,Siegel:1988gd,Metsaev:1995jp,%
  Angelopoulos:1997ij}.  By construction, these considerations concern
the gauge invariant spectrum of the theory. The second one is based on
equations of motion symmetries, i.e. on (quasi-)invariant differential
operators~\cite{Penrose:1965am,Eastwood:1987ki,Eastwood:2002su,Dobrev:2007vn}.
In particular, a technique to classify linear partial differential
equations for which Poincar\'e lifts to conformal symmetry was
developed in~\cite{Shaynkman:2004vu}.

Our considerations in this context will be restricted to free
classical (gauge) fields, i.e., to linear PDEs. So we will not address
any of the issues raised by the contemporary debate on scale versus
conformal invariance for interacting quantum field theories, see
e.g.~\cite{Nakayama:2013is,Dymarsky:2015jia} and references therein.

\subsection{Curvatures versus potentials}
\label{sec:curv-vers-potent}

Strictly speaking, the symmetry analysis described above applies to
PDEs without gauge symmetries. This is often sufficient because the
equations of motion of any linear gauge system admit a ``curvature''
formulation. A standard example consists of Fronsdal fields in (A)dS
or Minkowski spacetime which can be reformulated in terms of
gauge-invariant curvatures \cite{deWit:1979pe}. In the case of spin 1,
this is simply the formulation where the Faraday tensor $F_{\mu\nu}$
is the fundamental field. For spin $2$ (and higher), this is the
formulation in terms of the (generalized) Weyl tensor.

It is important to note that the formulation in terms of potentials
with gauge symmetries and the associated curvature formulation are not
equivalent when insisting on locality. In particular, they may have
different symmetries: for instance, at the level of equations of
motions, Fronsdal fields in $d=4$ with $s\geq 2$ are conformal in
terms of curvatures but not in terms of potentials. This is known to
experts but we are not aware of a detailed discussion in the
literature. In our approach, this is included by using the field-theoretic Batalin--Vilkoviski (BV) formalism, respectively the first quantized BRST approach
as described in the next sections, which allows us to provide a simple
proof in Section~\bref{sec:group-theor-obstr} below.

\subsection{Batalin-Vilkovisky formalism}
\label{sec:gauge-systems}

A better technical control on the degeneracies in Lagrangian gauge
systems has been achieved with the work of Batalin and Vilkovisky
\cite{Batalin:1981jr,Batalin:1983wj,Batalin:1984ss,Batalin:1985qj}
(see e.g. \cite{Henneaux:1992ig,Gomis:1995he} for reviews and
\cite{Barnich:1995db,Barnich:2000zw,Barnich:2010xq} for discussions in
the context of jet-bundles).

Let us denote by $\varphi^i$ the fields of the theory, by $x^\mu$ the
spacetime coordinates and by $\cL_0$ the Lagrangian. Under standard
regularity conditions, the notion of a generating set of gauge
generators $R^i_\alpha$ is crucial. Associated to a choice of such a
generating set, there is an extended set
$\{\phi^A\}=\{\varphi^i,C^\alpha,\cdots\}$ of fields $\varphi^i$,
ghosts $C^\alpha$, ghosts for ghosts, ... and their antifields
$\phi^*_A$, graded in terms of a ghost number and equipped with an
antibracket
\begin{equation}
  (\cdot,\cdot )=\int d^nx\,
  \vddr{\cdot}{\phi^A(x)}\vddll{\cdot}{\phi^*_A(x)}
  -(\phi\leftrightarrow
  \phi^*)\label{eq:37}.
\end{equation}
Furthermore, one can systematically construct a proper, minimal, ghost-number
$0$ solution
\begin{equation}
S=\int d^nx\,\big(\cL_0 +\varphi^*_i R^i_\alpha (C^\alpha)+\dots \big),\label{eq:18}
\end{equation}
to the Batalin-Vilkoviski master equation
\begin{equation}
  \half (S,S)=0\label{eq:24}. 
\end{equation}

\subsection{Local BRST cohomology}
\label{sec:local-brst-cohom-1}

Once the theory is reformulated within the BV formalism, a natural
question is the computation of local BRST cohomology, i.e., the
classification of the cohomology of the BRST differential
$s=(\cdot,S)$ in the space of local functionals. These groups do not
depend on the specific formulation of the theory, in the sense that
they can be shown to be invariant under the introduction/elimination
of (generalised) auxiliary fields \cite{Henneaux:1989ua}. In
particular:

{\em Equivalence classes of variational symmetries, up to on-shell
  vanishing variational symmetries and non trivial gauge symmetries
  with field dependent gauge parameters, are isomorphic to local BRST
  cohomology in ghost number $-1$. }

Whereas the computation of $H^{-1}(s)$ is in general rather involved,
the computation in lower ghost numbers is much easier. For instance,
in irreducible gauge systems for which the generating set of gauge
symmetries does not admit local degeneracies, one can show that there
is no cohomology in ghost numbers below $-2$, while cohomology in
ghost number $-2$ is given by equivalence classes of global
reducibility parameters, i.e., by sets of local functions $\bar
f^\alpha$ such that
\begin{equation}
R^i_\alpha(\bar f^\alpha)\approx 0,\label{eq:54}
\end{equation}
where $\approx$ means an equality on the surface defined by the
equations and their derivatives, with two sets of local functions
considered equivalent if they agree on this surface. 

\subsection{Constraints for variational symmetries}
\label{sec:antibr-map-glob}

The antibracket induces a well-defined bracket in local BRST
cohomology,
\begin{equation}
  \label{eq:48}
(\cdot,\cdot)_M:  H^{g_1}(s)\times H^{g_2}(s)\mapsto H^{g_1+g_2+1}(s). 
\end{equation}
When $g_1=-1=g_2$, it follows that $H^{-1}(s)$ is a (graded) Lie
algebra with respect to the above antibracket which is isomorphic, up
to a change of grading, to the Lie algebra of equivalence
classes of variational symmetries. Cohomology in fixed ghost number
$H^g(s)$ is a module thereof. In turn, this imposes constraints on
variational symmetries which we will use in our analysis below. More
precisely:

\begin{prop}
\label{prop1}
In the Lagrangian case, local BRST cohomology in the ghost number $g$,
$H^{g}(s)$, is necessarily a module of any subalgebra of $H^{-1}(s)$,
and thus of any subalgebra of the algebra of equivalence classes of
variational symmetries.
\end{prop}

Such a property can of course also be established without using the BV
formalism. For instance, that global reducibility parameters form a
module under variational symmetries has been shown directly in section
3.9 of \cite{Barnich:2001jy}. The reasoning can be summarized as
follows. In proper Lagrangian gauge systems, gauge symmetries form an
ideal in the set of all variational symmetries. This implies that,
on-shell, the commutator of a gauge symmetry with a global symmetry $\delta_X$
can be written in terms of the generating set,
\begin{equation}
  \label{eq:58}
  \delta_X R^i_\alpha(f^\alpha)-\delta_{R(f)} 
X^i\approx R^i_\alpha(X^\alpha_\beta(f^\beta)+\delta_X f^\alpha),
\end{equation}
for some total differential operators $X^\alpha_\beta$. When
$f^\alpha$ are reducibility parameters, both terms in the commutator
on the left hand side vanish on-shell. It follows that
$X^\alpha_\beta+\delta_X f^\alpha$ are also reducibility
parameters. One then proceeds to show that trivial variational
symmetries or trivial reducibility parameters are mapped to trivial
reducibility parameters. This implies that the module action is given
by
\begin{equation}
  \label{eq:60}
  \big([X^i],[\bar f]\big)\mapsto 
[X^\alpha_\beta(\bar f^\beta)+\delta_X \bar f^\alpha].
\end{equation}

In linear theories the generating set of gauge transformations can
usually be chosen to be field independent. If the same goes for the
reducibility parameters, like in the concrete example of Fronsdal
higher-spin fields in dimensions greater than 3 considered below, the
equation that determines the module action simplifies to
\begin{equation}
  \label{eq:75}
  -\delta_{R(f)} X^i\approx R^i_\alpha(X^\alpha_\beta(f^\beta)). 
\end{equation}

In the case where the linear gauge theory is the result of the
linearization of an interacting gauge theory around a solution
$\bar \phi$, the linear part of gauge transformations with gauge
parameters replaced by reducibility parameters,
$X^i=R^{i1}_\alpha(\bar f^\alpha)$, form a subalgebra of variational
symmetries. The module action is then described by a derived bracket
determined through the terms of the BRST extended cubic vertex of the
full interacting theory that contains the information on the gauge
algebra,
$%d^nx\, 
C^*_\alpha
C^\alpha_{\beta\gamma}[\bar\phi](C^\beta,C^\gamma)$.
This is discussed in detail in sections 4 and 7.4 of
\cite{Barnich:2001jy}. More generally, the derived bracket in the BV
formalism has been originally proposed in
\cite{Sen:1994ic,Nersessian:1995qb,Grigoriev:1998gn}.

\subsection{Linear Lagrangian theories}
\label{sec:linear-theories}

When the BRST differential $s$ is linear in the fields
$\Psi^\alpha=(\phi^A,\phi^*_A)$ of the BV formalism, it is determined
by a real ``first quantized BRST operator'' $\brst$ (see
\cite{Barnich:2004cr} and also \cite{Barnich:2010sw} for
conventions). Introducing an auxiliary superspace of wave functions
$\phi^\alpha(x)$ and basis elements $e_\alpha$ of opposite ghost
number and identical parity to $\Psi^\alpha$, and the string field
$\Psi=e_\alpha\Psi^\alpha$, the BRST operator is defined as
\begin{equation}
  \label{omega-s}
  s(\Psi)=\Omega(x,\d)\Psi,\quad
  \brst e_\alpha \phi^\alpha =e_\beta \brst^\beta_\alpha(x,\dl{x}) \phi^\alpha(x),
\end{equation} 
where $\d_\mu=\dl{x^\mu}+\Psi^\alpha_\mu\dl{\Psi^\alpha}+\dots$ denotes the
total derivative.  More generally, any linear differential operator of
the form $A^\alpha_\beta(x,\dl{x})$ acting from the left determines a
unique linear evolutionary vector field acting from the right and such
that
\begin{equation}
V_A (\Psi^\alpha)=A^\alpha_\beta
(x,\d)\Psi^\beta \label{eq:4}. 
\end{equation}
This map is one to one and moreover is a homomorphism, i.e.,
$\commut{V_A}{V_B}=V_{\commut{A}{B}}$.

In the variational case the space of fields is equipped with a
constant nondegenerate odd Poisson bivector
$\omega^{\alpha\beta}$. Its inverse
$\omega_{\alpha\gamma}\omega^{\gamma\beta}=\delta_\alpha^\beta$
satisfies
$\omega_{\alpha\beta}=(-1)^{1+\p{\alpha}\p{\beta}}\omega_{\beta\alpha}$
and determines an anti-symplectic form
\begin{equation}
  \omega(\phi,\chi)=\int d^dx\, \omega_{\alpha\beta} \phi^\alpha(x)
  \chi^\beta(x)\,, \qquad  
  \omega(\phi,\chi)=-(-)^{\p{\phi}\p{\chi}}\omega(\chi,\phi). 
\end{equation} 
An operator $A$ is called symplectic if 
\begin{equation}
\omega(A\phi,\chi)+(-)^{\p{A}\p{\phi}}\omega(\phi,A\chi)=0\,,
\end{equation}
where it is assumed that total derivatives do not contribute because
wave functions are assumed to vanish at infinity.

Symplectic operators are one to one with quadratic functionals,
\begin{equation}
  F_A=\half \int d^dx \, \Psi^\alpha
   \omega_{\alpha\beta} A^\beta_\gamma\Psi^\gamma.
\end{equation} 
The linear vector field $V_A$ associated to a symplectic operator $A$
is Hamiltonian, $V_A ( \Psi^\alpha)=\ab{\Psi^\alpha}{{F_A}}$, and satisfies
\begin{equation}
\label{homomorph}
  \ab{F_A}{F_B}=F_{\commut{A}{B}}\,. 
\end{equation} 
In particular, for linear theories, $\brst$ is symplectic and the
master action can be written as
\begin{eqnarray}
  \label{eq:16}
  S=F_\Omega\,,\qquad s=\ab{\cdot}{S}. 
\end{eqnarray}

Linear variational symmetries are determined by ghost number $-1$
quadratic functionals $\cK$ such that $(S,\cK)=0$. They are trivial if
$\cK=(S,\cT)$ with $\cT$ a quadratic ghost number $-2$ functional.
According to the above discussion, $\cK$ is determined by an even
ghost number $0$ symplectic operator $K$, while $\cT$ is odd ghost
number $-1$ symplectic operator $T$, and
\begin{equation}
  \label{eq:27}
  [\brst,K]=0,\quad K\sim K+[\brst,T].
\end{equation}
The problem of determining linear variational symmetries has thus been
rephrased as a problem of BRST operator cohomology. Furthermore, since
the Lie algebra structure of equivalence classes of linear variational
symmetries is encoded in the antibracket induced in local BRST
cohomology quadratic ghost number $-1$ functionals, it is also
represented by the commutator bracket induced in BRST operator
cohomology. It follows that:

{\em For linear proper gauge systems, there is thus an isomorphism
  between the Lie algebras of equivalence classes of linear
  variational symmetries and the commutator bracket of BRST operator
  cohomology in the space of symplectic ghost number $0$ operators.}

As a consequence, we have: 
\begin{prop}
\label{prop2}
BRST operator cohomology in the space of ghost number $g$ symplectic
operators $H^{g}_{\rm sym}([\brst,\cdot])$, is necessarily a module of
any subalgebra of $H^{0}_{\rm sym}([\brst,\cdot])$, and thus of any
subalgebra of the Lie algebra of equivalence classes of linear
variational symmetries.
\end{prop}

Suppose that one can prove that in a given ghost number $-g$,
representatives for local BRST cohomoloy can be chosen to be linear
functionals in the fields. This is for instance the case in ghost
number $-1$ for the Fronsdal fields discussed below. Without loss of
generality, they can always be chosen of the form
\begin{equation}
L_\phi=\int d^dx \,\Psi^\alpha\omega_{\alpha\beta} \phi^\beta(x) \label{eq:11}.
\end{equation}
It then follows from~\eqref{homomorph} 
that
\begin{equation}
 s (L_\phi)\equiv \ab{L_\phi}{S}=(-)^{1+\p{\phi}}L_\chi \quad \iff
 \quad \brst \phi  =\chi\,.
\end{equation} 
In particular, linear local BRST cohomology in ghost number $g-1$ is
isomorphic to BRST state cohomology in ghost number $g$, $H^g(\brst)$,
with vectors of the form $\phi=\phi^\alpha e_\alpha$.

Independently of this correspondence, we have:
\begin{prop}
\label{prop3}
BRST state cohomology in ghost number $g$, $H^g(\brst)$, is
necessarily a module of any sub-algebra of BRST cohomology for
symplectic, ghost number zero operators, $H^{0}_{\rm
  sym}([\brst,\cdot])$, and thus of any subalgebra of the Lie algebra
of equivalence classes of linear variational symmetries.
\end{prop}

Before making contact with a genuine first-quantized description, note
that in this work, the wave functions $\phi^\alpha$ and the associated
fields $\Psi^\alpha$ are taken real from the outset. By using the
parity automorphism $I^\alpha_\beta=(-1)^{|\alpha|}
\delta^\alpha_\beta$, the antisymplectic structure
$\omega_{\alpha\beta}$ determines an odd symmetric inner product,
\begin{equation}
\label{inner}
  \inner{\phi}{\chi}=\int d^dx (-1)^{\p{\alpha}}\omega_{\alpha\beta}
  \phi^\alpha(x)\chi^\beta(x) \,, 
\qquad
  \inner{\phi}{\chi}= ({-1})^{\p{\phi}\p{\chi}}\inner{\chi}{\phi}\,.
\end{equation} 
It turns out that Grassmann odd symplectic operators are formally
self-adjoint with respect to~\eqref{inner} while Grassmann even
symplectic operators are anti-self-adjoint. In particular, $\brst$ is
self-adjoint while representatives of global symmetries are
anti-self-adjoint. In concrete application it is often useful to work
in term of the symmetric inner product~\eqref{inner} in which case the
master action takes the form
\begin{equation}
S=\half \inner{\Psi }{-I\brst \Psi}\,.
\label{eq:9}
\end{equation}
Note that $-I\brst$ is also symplectic and self-adjoint. It is
equivalent to $\brst$ by a change of basis. When working with
\ref{inner}, we implicitly replace in what follows $-I\brst$ with
$\brst$ and similarly for the representatives of global symmetries,
so that the expression for the master action simply becomes $\half
\inner{\Psi }{\brst \Psi}$. This does not lead to confusions since
$H^\cdot(\brst)\simeq H^\cdot (-I\brst)$.

In a full quantum mechanical setting, one deals with a complex Hilbert
space and uses a complex structure such that $\omega_{\alpha\beta}$
becomes the imaginary part of the hermitean inner product. This type
of construction has been originally used in the context of string
field theory,
\cite{Bochicchio:1986bd,Bochicchio:1986zj,Thorn:1986qj,Thorn:1988hm}
(see also \cite{Barnich:2003wj} for an analysis from the point of view
of gauge systems).

\subsection{Constraints on equations of motion symmetries}
\label{sec:equat-moti-symm-1}

In the non-Lagrangian case, 
following~\cite{Barnich:2004cr,Barnich:2010sw,Barnich:2009jy}, we assume that 
the gauge system is described by a nilpotent, ghost number $1$ BRST differential 
$s$ represented by an evolutionary vector field on a bigraded jet-space of 
fields, which contains, besides the ghost number, an antifield 
number~\footnote{In fact the antifield number is determined by the ghost number and 
hence is not an independent ingredient of the definition.} according to which 
the BRST differential decomposes as $s =\delta+\gamma+s_1\dots$, with 
$\delta,\gamma,s_1,\dots$ of antifield number $-1$, $0$, $1$, $\dots$, such that 
the cohomology $\delta$ provides a homological resolution of the local functions 
defined on the surface determined by the original equations of motion. The 
differential $\gamma$ encodes the gauge symmetries of the equations of motion, 
which are required to close only on-shell, see 
e.g.~\cite{Henneaux:1989cz,Henneaux:1992ig} for more details.

When equipped with the commutator bracket, evolutionary vector fields 
form a graded Lie algebra, the bracket carrying degree $0$. Including
the BRST differential yields a differential graded Lie algebra, with
the bracket descending to cohomology,
\begin{equation}
  \label{eq:29}
  [\cdot,\cdot]_M: H^{g_1}([s,\cdot])\times H^{g_2}([s,\cdot])
  \longrightarrow H^{g_1+g_2}([s,\cdot]).
\end{equation}

In this context: 

{\em Equivalence classes of equations of motion symmetries modulo
  on-shell vanishing ones and non trivial gauge symmetries are
  described by the adjoint cohomology of $s$ in the space of
  evolutionary vector fields of degree $0$, $H^0([s,\cdot])$.}

Even though less restrictive than in the Lagrangian case, what we
will use to constrain equivalence classes of equations of motion
symmetries is:
\begin{prop}
\label{prop4}
The adjoint BRST cohomology $H^g([s,\cdot])$ is necessarily a module of
any subalgebra of $H^0([s,\cdot])$, and thus of any subalgebra of the
space of equivalence classes of equations of motion symmetries.
\end{prop}

Note that, in addition to equations of motion and Lagrangian systems,
one can consider a class of theories interpolating between these two
in the sense that the equations of motion are supplemented with a
Lagrange structure
\cite{Kazinski:2005eb,Kaparulin:2010ab,Kaparulin:2011aa}.

\subsection{Linear equations of motion}
\label{sec:line-equat-moti}

In the non-Lagrangian case, linear gauge systems are described by a
BRST operator $\brst$ that is no longer required to be symplectic. The
adjoint cohomology of $s$ in the space of evolutionary vector fields
that are linear in the fields and are of ghost number $g$ is
isomorphic to $H^g([\brst,\cdot])$, the adjoint BRST operator
cohomology in the space of operators of ghost number $g$. Hence,
equivalence classes of linear equations of motions symmetries are
described by adjoint BRST operator cohomology in degree $0$. The
analogs of Propositions \bref{prop2} and \bref{prop3} are then:

\begin{prop}
\label{prop5}
The adjoint BRST operator cohomology in ghost number $g$,
$H^g([\brst,\cdot])$, is necessarily a module of any subalgebra of
$H^0([\brst,\cdot])$, and thus of any subalgebra of the space of
equivalence classes of linear equations of motion symmetries.
\end{prop}

\begin{prop}
\label{prop6}
The BRST state cohomology in ghost number $g$, $H^g(\brst)$, is
necessarily a module of any subalgebra of $H^0([\brst,\cdot])$, and
thus of any subalgebra of the space of equivalence classes of linear
equations of motion symmetries.
\end{prop}

Note that, compared to the general case, inequivalent symmetries of
linear systems possess a richer structure. Namely, they form an
associative algebra with the product induced by the operator product
of cohomology representatives.

\subsection{Relation to the unfolded formalism}
\label{sec:unfolded}

Suitable modules under a spacetime symmetry algebra, closely related
to BRST state cohomology $H^{g}(\brst)$, play a crucial role in the
unfolded formulation of gauge field dynamics developed in the context
of higher-spin theories
\cite{Vasiliev:1988xc,Lopatin:1987hz,Alkalaev:2003qv,%
  Vasiliev:2005zu,Skvortsov:2008vs,Vasiliev:2009ck}. Typically, the
module $H^{-1}(\brst)$, or $H^{-p}(\brst)$ with maximal $p$ in
general, is an initial ingredient in terms of 1-form fields. The next
step consists in finding modules of $0$-form fields related to gauge
invariant curvatures such that the system of 1 and 0 forms is
consistent and gauge invariant.

The precise relation to the BRST first quantized formulation can be
understood by using the parent approach developed
in~\cite{Barnich:2004cr,Barnich:2006pc,Barnich:2010sw,%
  Grigoriev:2010ic,Grigoriev:2011gp}. Starting from a free gauge
system described by a nilpotent BRST operator $\brst$ as described in
Sections~\bref{sec:linear-theories} and \bref{sec:line-equat-moti},
the system is extended by allowing the wave functions to depend on
extra variables $y^\mu$, which are coordinates on the fibers of the
tangent bundle over spacetime, and Grassmann odd ghost variables
$\theta^\mu$, $\gh{\theta^\mu}=1$, to be identified with $dx^\mu$ and
associated to the constraints $(\dl{x^\mu}-\dl{y^\mu})\Phi=0$. The
parent BRST operator taking into account these new constraints along
with the original ones accounted in $\brst$ is
\begin{equation}
  \label{parent}
  \brst^P=\theta^\mu\left(\dl{x^\mu}-\dl{y^\mu}\right)+\bar\brst\,, 
  \qquad \bar\brst=\left.\brst\right|_{x^\mu\to x^\mu+y^\mu,
    \dl{x^\mu}\to \dl{y^\mu}}. 
\end{equation}
Note that~\eqref{parent} is a minimal version. In general, one can use
a generic parametrization of the tangent space and/or incorporate a
suitable (nonlinear) flat connection to account for specific
symmetries and/or spacetime geometries. Additional details can be
found in~\cite{Barnich:2006pc,Grigoriev:2006tt,Bekaert:2009fg}. Note
in particular that the associated field theories are related through
elimination of generalized auxiliary fields, provided the functional
space for $y^\mu$ is taken to be formal power series. This is not the
only interesting choice but the one that guarantees equivalence in the
sense of local field theories.

Consider the cohomology groups $\cH_p=H^{-p}(\bar\brst|_x)$ of the
second term in $\brst^P$ in the space of states at a given spacetime
point $x$. It is isomorphic to $H^{-p}(\brst)$ in the space of formal
power series at $x^\mu$. In applications, this space is often
isomorphic for all spacetime points $x$, as happens for instance if
the system has a symmetry group $G$ and is defined on a homogeneous
space of $G$. If in addition $\bar\brst$ can be made $x$-independent
by a $G$-transformation, i.e., $g\bar\brst g^{-1}$ does not depend on
$x$, this transformation makes the first term in~\eqref{parent} into a
$\mathfrak{g}$-covariant derivative in a specific representation
(see~\cite{Barnich:2006pc,Alkalaev:2008gi,Bekaert:2009fg,Alkalaev:2009vm}
for explicit examples and details).

If $\cH_p$ is $x$-independent, by eliminating generalized auxiliary
fields, the system can be reduced to an equivalent system whose states
take values in $\cH_p$ only. More precisely, dynamical fields (in
contrast to ghost, antifields, etc.) are $p$-forms $\phi_p$ with
values in $\cH_p$. The equations of motion and gauge symmetries for
the reduced system then have the following structure
  \begin{gather}
    \label{unf}
    (d+\sigma_1) \phi_0=0\,, \quad (d+\sigma_1) \phi_1+\sigma_2
    \phi_0=0\,, \quad
    (d+\sigma_1) \phi_2+\sigma_2 \phi_1+\sigma_3\phi_0=0\,,\quad
    \ldots\nonumber\\ 
    \delta \phi_0=0\,, \quad \delta \phi_1=(d+\sigma_1) \chi_1\,,
    \quad \delta \phi_2=(d+\sigma_1) \chi_2 +\sigma_2\chi_1 \,, \quad
    \ldots
\end{gather}
where $d=\theta\cdot\dl{x}$ and $\sigma_p$ are algebraic (i.e.,
$\dl{x}$-independent) operators 
 of order $p$ in $\theta$, and $\chi_k$
are gauge parameters which are $k-1$-forms with values in
$\cH_k$. Note that $d+\sigma_1+\sigma_2+\ldots$ is the homological
differential induced by $\brst^T$ in the cohomology of
$\bar\brst$. This is the minimal unfolded form of the equations and
the BRST state cohomology groups $H^{-p}(\brst)$ with a suitable
choice of the functional space are precisely the spaces of $p$-forms
in this formulation.

The cohomology $H^0(\bar\brst)$ is known in the unfolded approach as
the Weyl module. It consists of the gauge invariant (generalized) Weyl
tensor together with all of its on-shell inequivalent derivatives.
This space coincides with the space of gauge-inequivalent solutions
to the equations of motions in the space of formal power series.

Let us finally note that $H^{-p}(\bar\brst|_x)$ may in general differ
in distinct regions of spacetime and then it is not clear what the
minimal unfolded formulation is.  Typical examples are gauge fields
defined on the ambient space $\fR^{d+2}$ in the context of the
$AdS_{d+1}$/$CFT_d$ correspondence. For instance, in this case
$H^0(\bar\brst_x)$ for $x$ on the lightcone $X^2=0$ and on the
hyperboloid $X^2=\ell^2$ may well be different. It is this fact that
underlies the ambient space
approach~\cite{Bekaert:2012vt,Bekaert:2013zya} to boundary values of
$AdS$ gauge fields.

\subsection{Explicit construction of curvature formulations}
\label{sec:constr-curv-form}

Covariant curvature formulations can often be constructed directly
from group-theoretical arguments. For instance, such
formulations are well-known for $4$-dimensional Fronsdal fields
\cite{Bargmann:1948ck,Weinberg:1965rz} and Fradkin-Tseytlin fields
\cite{Fradkin:1985am}. For mixed-symmetry massless fields in Minkowski
spacetime, they can also be constructed in a direct way
\cite{Bekaert:2002dt,Bekaert:2003az,Bekaert:2015fwa}.

In principle, a systematic way to obtain a curvature formulation for a
given gauge system uses either the unfolded or the first-quantized
BRST approach. Indeed, from~\eqref{unf}, it follows that $\phi_0$ is
gauge invariant and that the equation for $\phi_0$ does not involve
other fields. This means that putting to zero all $\phi_l$ with $l>0$
gives a consistent unfolded system. This is the unfolded form of the
formulation in terms of curvatures as fundamental fields. The simplest
example is Maxwell's equations for the Faraday tensor.

This unfolded formulation of the curvature system is sometimes
difficult to construct. In all cases, a simple version of a curvature
formulation can be obtained from the parent system~\eqref{parent} by
putting to zero all fields which are forms of nonzero degree. More
precisely, the equations of motion and gauge symmetries then take the
form
\begin{equation}
\label{stuk}
  (\dl{x^\mu}-\dl{y^\mu})\Phi_0(x,y)=0\,, \quad \bar\brst
  \Phi_0(x,y)=0\,, \quad \delta_\chi \Phi_0=\bar\brst \chi\,, 
\end{equation} 
where $\gh{\Phi_0}=0$, $\gh{\chi}=-1$ and both $\Phi_0$ and $\chi$ are
$\theta^\mu$-independent. This system is equivalent to the above
unfolded formulation if one explicitly eliminates the pure gauge
degrees of freedom related to the algebraic gauge symmetries
in~\eqref{stuk}. It can thus be regarded as a St\"ueckelberg
description of the curvature system.

Hence, for general mixed-symmetry (partially)-massless fields in (A)dS
or Minkowski space, covariant curvature formulations are implicitly
contained in the unfolded or parent formulations constructed in
\cite{Alkalaev:2003qv,Skvortsov:2008sh,Skvortsov:2009zu%
  ,Boulanger:2008up,Boulanger:2008kw}, respectively
\cite{Alkalaev:2008gi,Alkalaev:2009vm,Alkalaev:2011zv}.

\section{Gauge fields in Minkowski spacetime}
\label{sec:conf-invar-fronsd}

\subsection{BRST formulation of Fronsdal fields}
\label{sec:massless-higher-spin}

The BRST formulation of higher-spin gauge fields
\cite{Ouvry:1986dv,Bengtsson:1987qb,Henneaux:1987cp,Sagnotti:2003qa}
can be summarized as follows. Take a $d$-dimensional Minkowski
spacetime with $d\geq 3$ in order to guarantee regularity assumptions
needed below and with metric $\eta_{\mu\nu}=\text{diag}
(-1,1\dots,1)$. The space of states is the Fock space of polynomials
in bosonic oscillators $\os^\mu$ (usually denoted by $a^\dagger_\mu$),
and fermionic ghosts $b,c$ with $\gh{c}=1, \gh{b}=-1$, tensored with
the space of functions in $x^\mu$ and the mass-shell ghost $c_0$,
$\gh{c_0}=1$.  The inner product is $\inner{\cdot}{\cdot}=\int
d^dx\, dc_0\, \inner{\cdot}{\cdot}_F$ where
$\inner{\cdot}{\cdot}_F$ is the standard inner product in the Fock
space for which $\os_\mu^\dagger=\dl{\os^\mu}$, $c^\dagger=\dl{b}$ and
$b^\dagger=-\dl{c}$.  The operators $x^\mu,c_0$ are self-adjoint
while $\dl{x^\mu}^\dagger=-\dl{x^\mu}$ and
$\dl{c_0}^\dagger=-\dl{c_0}$.

The self-adjoint BRST operator is 
\begin{equation} 
\begin{aligned}
\label{eq:Fbrst} 
& \Qbrst=c_0\,\Box+c\,\cS+\cS^\dagger
  \dl{b}+c\,\dl{b}\dl{c_0} ,
  \qquad \Qbrst^\dagger =\Qbrst,\\
&  \Box= \dl{x} \cdot \dl{x},\quad \cS=\dl{\os}\cdot\dl{x}\,, \quad
  \cS^\dagger =-\os \cdot \dl{x}\,.
\end{aligned}
\end{equation}
The ghost number operator is 
\begin{eqnarray}
\cG &=&c_0\dl{c_0} + c \,\dl{c} -b\, \dl{b},\qquad \cG^\dagger=1-\cG. 
\end{eqnarray}
Other operators that will be relevant are the BRST invariant
extensions of the trace $\cT$ and the level $N_s$,
\begin{equation}
\begin{aligned}
  & \cT=\dl{\os}\cdot \dl{\os}+2\,\dl{b}\dl{c}\,\\
  & N_s=\os \cdot \dl{\os}+c\,\dl{c}+b\, \dl{b}-s,\quad
  N^\dagger_0=N_0.
\end{aligned}
\end{equation}
The algebra satisfied by these operators is 
\begin{equation}
\begin{aligned}
  \label{eq:35}
&  [\Qbrst,N_s]=[\Qbrst,\cT]=[\cG,N_s]=[\cG,\cT]=0,\\
& [N_s,\cT]=-2\cT,\qquad [\cG,\Qbrst]=\Qbrst. 
\end{aligned}
\end{equation}
The string field is chosen as
\begin{multline}
  \label{eq:36}
  \Psi= \Big[ \Phi(x^\mu,\os^{\mu})+ c_0\,
  b\, B(x^\mu,\os^{\mu})+c\, b\,
  D(x^\mu,\os^{\mu}) + b\, C(x^\mu,\os^{\mu})+\\
-c_0\, \Phi^*(x^\mu,\os^{\mu})+ c\,
B^*(x^\mu,\os^{\mu})+
c_0\, c\, C^*(x^\mu,\os^{\mu})+c_0\, c\, b\, D^*
(x^\mu,\os^{\mu}) \Big] |0\rangle,  
\end{multline}
where the coefficients are expanded as power series in the oscillators
$\os^{\mu}$. The signs in the expansion have been choosen so that
the antibracket between a field and its antifield is $1$. The total
ghost number of the string field is $0$ and its parity is even. This
means that the ghost number and parities of the field coefficients are
opposite to those of the states.

We then have
\begin{multline}
  \label{eq:39}
  \Qbrst \Psi=\Big[c_0 \Box\Phi+c\, \cS \Phi+cc_0
b\, \cS B-c_0\, \cS^\dagger B +c\, B+c_0
 c b\, \Box D -c\, \cS^\dagger D +\\+c_0 b\, \Box
 C +c b\, \cS C +\cS^\dagger C - cc_0\, \cS
 \Phi^* + c_0 c\, \Box B^* +c_0 c\, \cS^\dagger
 D^*\Big]|0\rangle. 
\end{multline} 
The classical action for a spin $s\geq 0$ field is
\begin{equation}
\begin{aligned}
  \label{eq:3}
 & S[\Psi^T_{0,s}]=\half\langle \Psi^T_{0,s},\Qbrst \Psi^T_{0,s}\rangle,\\
 & \cT\Psi^T_{0,s}=0,  \quad N_s \Psi^T_{0,s}=0, \quad \cG \Psi^T_{0,s}=0,
\end{aligned}
\end{equation}
while the Batalin-Vilkovisky master action is 
\begin{equation}
\begin{aligned}
  \label{eq:38}
  & S[\Psi^T_{s}]=\half\langle \Psi^T_{s},\Qbrst \Psi^T_{s}\rangle, 
\\ & \cT\Psi^T_{s}=0,  \quad N_s \Psi^T_{s}=0.
\end{aligned}
\end{equation}
For $d=4$, action \eqref{eq:3} coincides, up to auxiliary fields, with
the gauge theory for free massless fields of helicity $\pm s$
introduced by Fronsdal \cite{Fronsdal:1978rb}. 

Explicitly, by doing the ghost inner product,
\begin{multline}
  \label{eq:40}
  S[\Psi^T_{s}]=\half\int d^dx\,\big[\langle \Phi, \Box \Phi\rangle_F
  -\langle D,\Box D\rangle_F -2\langle B,\cS \Phi\rangle_F
    +2\langle B,\cS^\dagger D\rangle_F  -\langle B,B\rangle_F\\
-2 \langle\Phi^*,\cS^+ C\rangle_F 
-2\langle D^*, \cS C\rangle_F  -2\langle B^*, \Box C\rangle_F \big].
\end{multline}

Removing the level $N_s$ constraint gives the sum of the free (master)
actions for all integer massless spins, while removing the trace
constraint $\cal T$ at fixed spin $s$ gives a model that contains, for $d=4$,
massless fields with helicities $-s,-s+2,\dots,s-2,s$ (see
e.g.~\cite{Barnich:2005ga} for a proof in the current context).

Finally, in order to explicitly deal with the trace constraint for
Fronsdal fields, we need:
\begin{prop}\label{propsymgen-alt}
  The Lie algebra of (anti-self-adjoint) operators defined on $\Ker
  \cT$ can be described by operators $A$ such that 
\begin{equation}
\label{symdef}
 \cT A=B \cT\,,\qquad A\sim A+C\cT\,,
\end{equation} 
where $B$ and $C$ are some operators (such that $A$ and $C\cT$ are
anti-self-adjoint).
\end{prop}
The statement is equivalent to the regularity of the equation
$\cT\phi=0$ or, more precisely, that any operator $C$ such that
$C\phi=0$ for all $\phi\in \Ker \cT$ can be written as $C=B\cT$ for
some operator $B$.  To see this, note that the Lie algebra spanned by
$\cT,\cT^\dagger,N_{1-\frac{\,\,d}{\,\,2}}$ is isomorphic to
$\mathfrak{sl}(2,\mathbb R)$, which is clear from the identification
$E_+:=\frac{\,1}{\,2}\cT^\dagger$, $E_-:=-\frac{\,1}{\,2}\cT$ and
$H:=N_{1-\frac{\,\,d}{\,\,2}}\,$.  Therefore any element in the
representation space has a unique decomposition
$\Phi=\phi_0+\cT^\dagger \phi_1+(\cT^\dagger)^2\phi_2+\ldots$ where
$\cT\phi_l=0$. Moreover, the projector $\Pi$ to the subspace $\Ker\cT$
of elements satisfying $\cT\chi=0$ can be written as
$\Pi=1-\cT^\dagger\Theta \cT$ for some
$\Theta(H,\cT,\cT^\dagger)$. Note that $\Ker\cT$ is orthogonal to $\im
\cT^\dagger$ and $\Pi$ is self-adjoint. Regularity then follows from
the structure of the projector $\Pi$. Indeed, $C\phi=0 \,\,\forall
\phi \in \Ker \cT$ implies $C\Pi=0$ which in turn gives
$C=C\cT^\dagger\Theta\cT$.

The space of operators on $\Ker \cT$ can be identified with the
quotient space of operators preserving $\Ker \cT$, i.e., $\cT
A\phi=0\, \forall \phi\in \Ker \cT$, modulo operators that act
trivially, i.e., $A\phi=0\, \forall \phi\in \Ker \cT$. Thanks to the
regularity of $\cT$, this space can be written as~\eqref{symdef} with
$B=\cT A\cT^\dagger\Theta$. If one is interested in anti-self-adjoint
operators, it is enough to require both $A$ and $C\cT$ to be
anti-self-adjoint. This completes the proof.

Note that \eqref{symdef} is the usual definition of the space of
inequivalent linear symmetries of the equation $\cT \phi=0$. The above
proof applies equally well to the Klein--Gordon equation
$\Box\varphi=0$ because $\Box$ enters an $sl(2,\fR)$-algebra together
with operators $x^2, x\cdot \dl{x}+\frac{d}{2}$ so that \eqref{symdef}
with $\cT$ replaced by $\Box$ coincides with the definition of linear
symmetries for the Klein--Gordon equation~\cite{Eastwood:2002su},
discussed here in
Section~\bref{sec:KG}. Proposition~\bref{propsymgen-alt}, with $\cT$
or $\Box$, is the first-quantized version of the acyclicity of the
associated Koszul differential in the field-theoretical picture.

For our purpose below, it is convenient to characterize operators on
$\Ker \cT$ differently. Any operator $A$ on the entire representation
space determines an operator $\Pi A\Pi$ on $\Ker \cT$. Conversely, an
operator on $\Ker \cT$ can be lifted to the entire space. Using the
expression for the projector, one finds that trivial operators on
$\Ker \cT$, i.e., those satisfying $\Pi A \Pi=0$ are of the form
$A=\cT^\dagger \alpha + \beta \cT$ for some operators
$\alpha,\beta$. It follows that operators on $\Ker \cT$ can be
described as the quotient space of all operators modulo those of the
form $\cT^\dagger \alpha + \beta \cT$. In particular,
anti-self-adjoint operators are described by the following quotient
\begin{equation}
 \label{propsymgen}
 A\sim A+\cT^\dagger \gamma - \gamma^\dagger\cT\,,
\end{equation} 
for some operator $\gamma$. It is important to note that this quotient
is only compatible with the commutator, i.e., operators
$A^\prime=\cT^\dagger \gamma - \gamma^\dagger\cT$ equivalent to zero
form an ideal in the Lie algebra of anti-self-adjoint operators if one
restricts oneselves in addition to operators that preserve $\Ker \cT$,
i.e. $\cT A^\prime=\delta \cT$ for some $\delta$. Indeed, $\cT A^\prime=\delta
\cT$ implies $\cT \cT^\dagger
\gamma=(\delta-\cT\gamma^\dagger)\cT$. Applying $\cT^\dagger\Theta$ to
both sides and using $\Pi \cT^\dagger=0$ one finds $\cT^\dagger
\gamma=\cT^\dagger\Theta (\delta-\cT\gamma^\dagger)\cT$, so that
$A^\prime=B\cT$ for some $B$. So if we restrict to operators
preserving $\Ker \cT$ then those of the form $\cT^\dagger \gamma -
\gamma^\dagger\cT$ form an ideal identical to the one in the
proposition~\bref{propsymgen-alt}.

\subsection{Classification of variational symmetries of a massless
  scalar}\label{sec:KG}

Let us now concentrate on a massless scalar, for which $s=0$ in the
above description, and use the existing classification of symmetries
of the equations of motion \cite{Eastwood:2002su} to infer the
classification of variational symmetries.  

In this case, the BRST operator reduces to
\begin{equation}
 \Qbrst=c_0\Box\,,
\end{equation} 
while the general expression for a ghost number $0$ or $-1$ operators is
\begin{equation}
 \sA=A(x,\dl{x})+B(x,\dl{x})c_0\dl{c_0},\quad
 \sD=D(x,\dl{x})\dl{c_0}. 
\end{equation} 
The condition that $\sA$ represents an element of $H^{0}([\Qbrst,\cdot])$,
\begin{equation}
\begin{gathered}
\label{ABeom}
\commut{\Qbrst}{\sA}=0\,,\qquad \sA\sim \sA+\commut{\Qbrst}{\sD} \quad  \\
\iff \quad \commut{\Box}{A}-B\Box=0\,, \quad A \sim
A+D\Box,\quad  B\sim B+\commut{\Box}{D},
\end{gathered}
\end{equation} 
coincides with the definition of linear symmetries used in
\cite{Eastwood:2002su}. 

The linear space $H^0(\commut{\Qbrst}{\cdot})$ of inequivalent linear
symmetries of the equations of motion (EOM) is an associative algebra
$\algA$ also known as higher-spin algebra~\cite{Vasiliev:2003ev}. For
a given symmetry $\sA$ let $A_S(x,p)$, $B_S(x,p)$ determine its
principal symbols, e.g. $A_S$ is the highest derivative term in $A$
where $\dl{x^\mu}$ is replaced with the commuting variable $p_\mu$. It
was shown in \cite{Eastwood:2002su} that \eqref{ABeom} implies that
\begin{equation}
 p\cdot \dl{x}A_S(x,p)=p^2 B_S(x,p)
\end{equation} 
i.e., that $A_S$ is a conformal Killing tensor and also that
inequivalent linear EOM symmetries are uniquely determined by their
principal symbols. It follows that as a linear space, $\algA$ is
isomorphic to the space of conformal Killing tensors.

Let us now turn to the space of linear, inequivalent, variational
symmetries. According to the general considerations above, they are
described by $H^0(\commut{\Qbrst}{\cdot})$ in the space of
(anti)-self-adjoint operators. Explicitly,
\begin{equation}
  \sA+\sA^\dagger=0 \iff  A+A^\dagger+B=0\,.\label{ABvar}
\end{equation} 
It is instructive to check that~\eqref{ABeom} and \eqref{ABvar} imply
that $\delta \phi=A\phi$ is indeed a linear variational symmetry
associated to
\begin{equation}
 S=\half \int d^dx\, \phi \Box\phi\,.
\end{equation} 

The elements from $H^0(\commut{\Qbrst}{\cdot})$ satisfying
$\sA=-\sA^\dagger$ form a Lie, but not an associative, subalgebra of
$\algA$. In this case, \eqref{ABvar} implies that
$A_S(x,-p)=-A_S(x,p)$ and $B_S(x,-p)=B_S(x,p)$. In other words: 

{\em For a massless real scalar, $H^{0}_{\rm sym}([\brst,\cdot])$, the
  space of inequivalent linear variational symmetries, is isomorphic
  to the space of conformal Killing tensors of odd rank.}

\subsection{Poincar\'e and dilatation symmetries of Fronsdal fields}
\label{Poincaredilatation}

Consider a real spacetime vector field $\xi(x)$ 
and the anti-self-adjoint, even, ghost number $0$ generator
\begin{equation}
\begin{split}
  \label{eq:1}
  & \CG=-\big(\xi \cdot \d+\half
  S^{\mu\nu}\Sigma_{\mu\nu}+\frac{\d\cdot
    \xi}{d}\Delta+\frac{\d^\mu(\d\cdot\xi)}{2d}\kappa_\mu\big),\\
  & S^{\mu\nu}=\half(\d^\nu\xi^\mu-\d^\mu\xi^\nu), \quad
  \Sigma_{\mu\nu} =\os_\nu\dl{\os^\mu}-\os_\mu\dl{\os^\nu},\\
  & \Delta=\frac{d}{2}-1+2c_0\dl{c_0} +c\dl{c}-b\dl{b},\quad
  \kappa_\mu= 4c_0\left(\dl{\os^\mu} b+\os_\mu \dl{c}\right),
\end{split}
\end{equation}
satisfying
\begin{equation}
  \label{eq:8}
  [\cT,\CG]=0= [N_s, \CG]. 
\end{equation}
By direct computation, one finds 
\begin{equation}
  \label{eq:6}
  [\Qbrst,\CG] = 2\beta^\lambda\Big((c \dl{\os^\lambda}+
  \os_\lambda \dl{b})N_{3-\frac{d}{2}}-c
  \os_\lambda\cT-\cT^\dagger \dl{\os^\lambda} \dl{b}\Big).
\end{equation}
if $\xi$ describes infinitesimal conformal
transformations, 
\begin{equation}
  \label{eq:2}
 \d_\mu\xi_\nu+\d_\nu\xi_\mu=
\frac{2}{d}\eta_{\mu\nu}\d\cdot\xi \iff \xi_\mu=a_\mu+\omega_{[\mu\nu]}x^\nu+\alpha
x_\mu+2x_\mu \beta\cdot x-\beta_\mu x\cdot x, 
\end{equation}
with constant parameters $a_\mu,\omega_{[\mu\nu]},\alpha,\beta_\mu$. 

So, the form of the operator $\Xi$ in \eqref{eq:1} has been fixed by
the following requirements: (i) It starts with $-\xi\partial$
implementing the spacetime transformations, (ii) it is antihermitian
and (iii) its commutator with the BRST operator produces either zero
or, at worst, a term which does not depend on the spacetime operators
$x^\mu,\dl{x^\mu}$. By themselves, these requirements have lead to
conformal vector fields. The conformal transformations and algebra are
thus recovered from this construction.

In the current framework, this confirms that infinitesimal Poincar\'e
and dilatation transformations, for which $\beta_\mu=0$, are
variational symmetries of Fronsdal's higher-spin gauge theory in all
dimensions $d\geq 3$. Furthermore, for the Klein-Gordon action, for
which $s=0$, the same holds for infinitesimal special conformal
transformations, since $\langle \Psi^T_{0,0},[\Qbrst,\CG]
\Psi^T_{0,0}\rangle=0$. The explicit form of the generators
$P_\mu,M_{\mu\nu},-D,K_\mu$ are obtained by differentiating $\CG$ with
respect to the parameters.

The last two terms in \eqref{eq:6} do not contribute for a spin $s$
field because its master action is $S=\half
\langle\Psi^T_s,\Qbrst\Psi^T_s\rangle$ with $\cT \Psi^T_s=0$ and
$N_s\Psi^T_s=0$.  Defining $\cK=\half \langle
\Psi^T_s,I\CG\Psi^T_s\rangle$ we thus get,
\begin{eqnarray}
  \label{eq:7}
  (S,\cK)=\half \langle\Psi^T_s,[\Qbrst, \CG]\Psi^T_s\rangle=
  \beta^\lambda\left(s-3+\frac{d}{2}\right)\langle\Psi^T_s, (c\dl{\os^\lambda}+
  \os_{\lambda} \dl{b}) \Psi^T_s\rangle.
\end{eqnarray}

It follows that:

{\em For spin $0$, there is conformal invariance at the level of the
  action in any dimension. For spin $s= 1$, this is the case for
  $d=4$.}

In the next section, we will first use the strategy outlined in
Section~\bref{sec:generalities} to quickly show that: 

{\em For $d\geq 3$, Fronsdal fields with $s\geq 2$ are invariant
  under Poincar\'e transformations and dilatations, but not
  conformally invariant, neither at the level of the action, not at the
level of the equations of motion}. 

We will then provide a direct proof that $\Xi$ cannot be modified so
as to include special conformal transformations among the variational
symmetries, i.e., among the generators commuting with $\Qbrst$.

\subsection{Obstructions to special conformal symmetries for Fronsdal
  fields}
\label{sec:group-theor-obstr}

\subsubsection{Obstructions at the level of the action}

Local BRST cohomology in ghost number $-2$, $H^{-2}(s)$, corresponds
in the current conventions to BRST state cohomology in ghost number
$-1$, $H^{-1}(\brst)$, and has been worked out in
\cite{Barnich:2005bn,Barnich:2004cr}. This space manifestly enters the
unfolded formulation of Fronsdal fields as the module of 1-form fields
and has originally appeared in this context
in~\cite{Lopatin:1987hz}. For a given spin $s$ gauge field, it is
represented by the vector space $V$ of elements of the form
\begin{equation}
  \label{eq:14}
  b\, A(x,\os)=b\, \os^{\mu_1}\dots \os^{\mu_{s-1}}\sum_{m=0}^{s-1}
  A_{\mu_1\dots\mu_{s-1}|\nu_1\dots \nu_m} x^{\nu_1}\dots
  x^{\nu_m}, 
\end{equation}
where $A(x,a)$ satisfies
\begin{equation}
\label{H-1}
 \left(\os\cdot \dl{x}\right)A=%0\,, \qquad
 \left(\dl{\os}\cdot\dl{x}\right)A
=\left(\dl{\os}\cdot\dl{\os}\right)A=\left(\dl{x}\cdot\dl{x}\right)A=0\,, 
\end{equation} 
and describes rank $s-1$ traceless Killing tensor fields on Minkowski
spacetime. The coefficients $A_{\mu_1\dots\mu_{s-1}|\nu_1\dots \nu_m}$
are totally traceless and have the symmetries of two-row Young
tableaux.

According to Proposition \bref{prop3}, the vector space $V$ is a
module for Poincar\'e and dilatation transformations, $\CG V\subset V$
when $\beta_\mu=0$ and we will work out the constraints coming from the
condition that $V$ be a module under special conformal transformations
as well by using standard representation-theoretic arguments.

The subspace $V_0 \subset V$ annihilated by the translation generators
$P_\nu=-\ddl{}{x^\nu}$ is
\begin{equation}
  b\,\os^{\mu_1}\dots \os^{ \mu_{s-1}}
  A_{\mu_1\dots\mu_{s-1}},\label{eq:30}
\end{equation}
with symmetric traceless constant tensors
$A_{\mu_1\dots\mu_{s-1}}$. The subspace $V_0$ is an irreducible
$o(d-1,1)$ (i.e. Lorentz) module.

Let us first assume that the action on $V$ of the Poincar\'e algebra
extended by dilatations lifts to $o(d,2)$ by including the special
conformal generators $K_\mu$. Using the explicit form of the
dilatation generator gives $Dv=(x\cdot\dl{x}+\frac{d}{2}-2)v$ for
$v\in V$. It follows from~\eqref{eq:14} that the spectrum of the
dilatation generator is given by $\frac{d}{2}-2,\frac{d}{2}-1,\ldots,
\frac{d}{2}-3+s$. At the same time, $D$ can be taken as a generator of
an $sl(2)$ subalgebra in $o(d,2)$, formed by $D,P_1,K_1$ say. It
follows that, in any finite-dimensional module, its spectrum must be
symmetric with respect to $0$. This shows that, for $d\geq 4$ and
$s>0$, the only option is $d=4,s=1$, which is indeed
conformal. Formally, in lower dimensions there are extra
possibilities: $d=2,s=3$ and $d=3,s=2$.  The former does not work
because $dim(V)=2$ and there is no 2-dimensional $sl(2)$ irreducible
representation with weights $-1,0,+1$.  The latter is ruled out as all
weights of $o(d,2)$ must be simultaneously either integer or
half-integer but $s$ is an integer while the eigenvalues of $D$ are
$\pm 1/2$.

\subsubsection{Obstructions at the
  level of equations of motion}

If we are only interested in equations of motion symmetries, the value
of the lowest weight $\Delta_L$ of the dilatation operator is not
known a priori and an extra analysis is needed. Indeed, in the
analysis above, this weight was fixed from the requirement that the
symmetry generator needed to be anti-self-adjoint. 

Let us restrict ourselves to $d\geq 3$. Any $o(d,2)$-module having
$V_0$ as a Lorentz sub-module annihilated by all translation
generators $P_\nu$, and hence lowest-weight with respect to
dilatations, can be induced from $V_0$ in a standard way: first pick
$\Delta_L$, which must be constant on $V_0$ because $V_0$ is Lorentz
irreducible and dilatation generators commute with the Lorentz
subalgebra, and then consider the (generalized) Verma module generated
from $V_0$, i.e., consider all formal combinations
$K_{\lambda_1}\ldots K_{\lambda_m} v$ where $v\in V_0$. Any
$o(d,2)$-module containing $V_0$ as a Lorentz submodule and such that
the translation generators $P_\nu$ annihilate $V_0$ is by construction
a quotient of this Verma module, as a consequence of the universality
property of Verma modules. Moreover, $\Delta_L$ must take special
values in order for the quotient to be finite-dimensional. More
precisely, $-\Delta_L$ has to be integer and such that $-\Delta_L \geq
s-1$. In other words the highest-weight $(-\Delta_L,s-1)$ must be integral
dominant. The corresponding finite-dimensional $o(d,2)$-module is
described by a two-row Young tableau (YT) with first row of length
$-\Delta_L$ and second row of length $s-1$, which will be written
$(-\Delta_L,s-1)$. Already for $-\Delta_L = s-1$, the spectrum of the
dilatation generator contains all integers from $1-s$ to $s-1$ and
hence at least $2s-1$ irreducible Lorentz components. However, the
vector space $V$ spanned by elements of the form \eqref{eq:14} instead
contains $s$ irreducible Lorentz components. For $-\Delta_L > s-1$, the
finite-dimensional $o(d,2)$-modules with highest-weight
$(-\Delta_L,s-1)$ contain even more than $2s-1$ irreducible Lorentz
components. Therefore, the only possibility is the trivial
representation: $s-1=\Delta_L=0$.

Another way to see that these modules cannot coincide is to observe
that the $o(d,2)$-module associated with the YT $(s-1,s-1)$ is the one
of conformal Killing tensor fields of rank $s-1$ in $d$
dimensions. The latter cannot coincide with the Poincar\'e module of
usual Killing tensor fields unless it is trivial, i.e., unless
$s=1$. In this way, we conclude that Fronsdal fields do not admit
special conformal transformations as equations of motion symmetries
unless $s=0,1$.

To see that for $s=1$, conformal symmetry is present for $d=4$ only,
the argument based on $H^{-1}(\brst)$ is not enough and $H^0(\brst)$
needs to be analyzed. It is well-known that the space of inequivalent
solutions to Maxwell equations is not conformal unless
$d=4$\footnote{See for instance, \cite{Shaynkman:2004vu} where
  conformal equations were classified by listing all suitable
  conformal modules. In the present language, the cohomology
  $H^0(\brst)$ is evaluated in the space of formal power series in
  $x^\mu$ in terms of generalized Verma modules.}. This implies that
Fronsdal fields in terms of potentials do not admit conformal symmetry
at the level of equations of motion, unless $s=0$ or $s=1,d=4$.

To conclude the discussion of Fronsdal fields in $d=4$, note that, as
a linear space, $H^{-1}(\brst)$ can be made into an
$o(4,2)$-module. This does not, however, correspond to an extension of
the Poincar\'e symmetries in the realisation of
Subsection~\bref{Poincaredilatation} and, moreover, it works only for
the complexified module because an (anti)self-duality condition should
be imposed. The idea is to start with the contragredient module
structure on the same linear space $V$ defined in~\eqref{H-1}. For
instance, introducing the standard inner product on polynomials, i.e.,
the one determined by $\inner{1}{1}=1$, $x_\mu^\dagger=\dl{x^\mu},
(\os_\mu)^\dagger=\dl{\os^\mu}$ so that for instance
$\inner{x^\mu}{x^\nu}=\eta^{\mu\nu}$, and defining new Poincar\'e
generators through $P^\prime_\mu=-P^\dagger_\mu$ and
$M_{\mu\nu}^\prime=-M_{\mu\nu}^\dagger$, one finds that the subspace
annihilated by $P^\prime_\mu$ is precisely the Lorentz-module
associated to the YT $(s-1,s-1)$. Taking $-\Delta_L=s-1$ one finds
that, as a complex module, $H^{-1}({\brst})$ lifts to an
$o(4,2)$-module described by the YT $(s-1,s-1,s-1)$. Details of
non-branching for this module can be found in
Section~\bref{sec:branching-rules}.

\subsubsection{Direct obstructions to special conformal generators}
\label{sec:direct-obstructions}

Let us now complete the analysis started in
Section~\bref{Poincaredilatation} and show directly that one cannot
modify $\Xi$ in \eqref{eq:1} so as to include special conformal
transformations among the variational symmetries when $d\geq 3$ and
$s\geq 2$.

Comparing to equation (II.10) of \cite{Mack:1969rr}, all the spacetime
dependence of the special conformal transformations is correctly
reproduced by $\CG$. It then follows from the analysis in this
reference that the only freedom left is to add a spacetime independent
operator linear in $\beta_\lambda$, or more precisely, to change
$\kappa_\mu$ to $\tilde \kappa_\mu=\kappa_\mu+\kappa'_\mu$ by the
addition of a $x^\mu,\dl{x^\mu}$-independent operator $\kappa'_\mu$
such that
\begin{equation}
  [\Delta, \kappa'_\mu]=\kappa'_\mu,\
  [\Sigma_{\mu\nu},\kappa'_\lambda]=\eta_{\mu\lambda}\kappa'_\nu-
  \eta_{\nu\lambda}\kappa'_\mu,\
  [\kappa_\mu,\kappa'_\nu]+[\kappa'_\mu,\kappa_\nu]+
  [\kappa'_\mu,\kappa'_\nu]=0\label{eq:5}.   
\end{equation}
We thus want to show that no such modification allows one to remove the
obstruction proportional to $\beta^\lambda$ on the right hand side of
\eqref{eq:7}.

Using Proposition \bref{propsymgen-alt}, formulated as in
~\eqref{propsymgen}, a symmetry generator $K$ needs to satisfy
$\cT K=B\cT$ and
$[\brst,K]=\cT^\dagger\gamma-\gamma^\dagger\cT\,$. Combining the
ansatz $K=\Xi+\beta^\mu\kappa'_\mu$ with equation \eqref{eq:6}, the
no-go result is proven if one can show that there does not exist an
operator $\kappa'_\mu$ independent of $x,\dl{x}$ satisfying
\eqref{eq:5} such that
\begin{equation}
  \label{eq:kappa}
  [\brst,{\kappa'}^{\lambda}]=-2 (c \dl{a_\lambda}+
a^\lambda \dl{b})N_{3-\frac{d}{2}}-
A^{\dagger\lambda}\cT+\cT^\dagger A^{\lambda}.
\end{equation}
for some operators $A^\lambda$.

First, using a decomposition according to the degree of homogeneity in
$x^\mu$, one can take without loss of generality in \eqref{eq:kappa}
that $\brst$ reduces to $c\dl{b}\dl{c_0}$ and that $A^\lambda$ is
$x,\dl{x}$ independent.

Second, decomposing operators $A=\sum_n A_n$ according to the level
associated to $N_0$, one gets in degree $0$, 
\begin{equation}
  \label{eq:kappan}
 [c\dl{b}\dl{c_0},{\kappa'}^\lambda_0]=-2(c \dl{a_\lambda}+a^{\lambda}
  \dl{b})N_{3-\frac{d}{2}}- A^{\dagger\lambda}_{2}\cT+\cT^\dagger
  A^{\lambda}_{-2}\,.
\end{equation}
Only this equation is relevant since at level different from zero, the
first term on the right hand side does not contribute and one can
choose the trivial solution
${\kappa'}^\lambda_n=0=A^{\lambda}_{n-2}=A^{\dagger\lambda}_{n+2}$.

Third, using the Lorentz transformation properties, one can assume that
${\kappa'}^\lambda_0=f_1\dl{a_\lambda}+a^\lambda g_{-1}$, where $f,g$
depend only on Lorentz invariant combinations of $a,\dl{a}$, or, by
suitably completing these invariants, that
$f_1=f_1(\cT^\dagger,N_{s},\cT,c_0,\dl{c_0},c,\dl{c},b,\dl{b})$
and similarly for $g_{-1}$. Since $c\dl{b}\dl{c_0}$ commutes with
$N_{s}$, one can restrict to the zero eigenspace of $N_s$, which means
in particular that one 
considers a theory at fixed spin $s$. Writing all operators in
normal-ordered form 
with respect to $\cT,\cT^\dagger$, i.e. in the form
$B=\sum_{l,m}(\cT^\dagger)^l \alpha_{lm} (\cT)^m$ for some
$\cT,\cT^\dagger$-independent $\alpha_{lm}$, and 
using
$\commut{c\dl{b}\dl{c_0}}{\cT}=\commut{c\dl{b}\dl{c_0}}{\cT^\dagger}=0$
the lowest order equation gives 
\begin{equation}
  \label{eq:15}
  [c\dl{b}\dl{c_0},{\kappa'}^\lambda_0]=-2(s-3+\frac{d}{2})(c
  \dl{a_\lambda}+a^{\lambda} 
  \dl{b}). 
\end{equation}

Fourth, decomposing $f_1=f_1^0+c_0 f_1^1$, where $f_1^0$ does not depend on
$c_0$, and similarly for $g_{-1}$, the equation implies 
\begin{equation}
  \label{eq:17}
  c\dl{b}(f^1_1\dl{a_\lambda}+a^\lambda
  g^1_{-1})=-2(s-3+\frac{d}{2})(c\dl{a_\lambda}+a^\lambda\dl{b}). 
\end{equation}
Finally, equating $c$-independent terms it follows that $s-3+\frac{d}{2}$ has to
vanish, which is only possible for $s=1$ and $d=4$, and for $s=2,d=2$
which is excluded from the discussion.

\subsection{Generic massless bosonic fields in Minkowski spacetime}

Mixed-symmetry massless fields were originally described
in~\cite{Labastida:1986gy,Labastida:1989kw} while further developments
relevant in the present context can be found in
\cite{Burdik:2001hj,Sagnotti:2003qa,Skvortsov:2008sh}, and also in
\cite{Alkalaev:2008gi} which we follow below. These systems are
variational and admit a Lagrangian formulation based on a BRST
operator $\brst$ generalizing the first quantized description of
Fronsdal fields reviewed in Section~\bref{sec:line-equat-moti}.

In $d$-dimensional Minkowski spacetime, generic mixed-symmetry
massless bosonic field of spin $s_1,\,\ldots,s_p$, the weights of the
respective little group representation, and where the number of rows
satisfies $p\leq [\frac{d-2}{2}]$, $[a]$ denotes the integer part of
$a\in\mathbb R$, can be described by the equations
\begin{gather}
\label{TT}
\dl{\os_i}\cdot \dl{\os_j}\Phi=0\,, \qquad \dl{\os_i}\cdot
\dl{x}\Phi=0\,, \qquad  \dl{x}\cdot \dl{x}\Phi=0\,,\\
\label{YC}
\os_i\cdot \dl{\os_j}\Phi=0\quad i > j\,, \qquad (\os_i\cdot
\dl{\os_i}-s_a)\Phi=0\,,  
\end{gather}
where we use, as usual, variables $\os^\mu_i$ with $\mu=0,\ldots, d-1$
and $i=1,\ldots,p$ to contract indices and work in terms of a
generating function $\Phi$.

In terms of the generating function $\Phi$
the gauge transformations read as
\begin{equation}
 \delta \Phi=\Qred\chi^{(1)}\,, \qquad \Qred=\left(\os_i \cdot
   \dl{x}\right)\dl{b_i}\,, 
\end{equation} 
where $\chi^{(1)}=b^i\chi^{(1)}_i(x,\os)$. For convenience, we
introduced here Grassmann-odd ghost variables $b^i$. The same operator
$\Qred$ determines gauge for gauge symmetries $\delta
\chi^{(1)}=Q\chi^{(2)}$ etc. Gauge (for gauge) parameters satisfy the
analog of \eqref{TT} and the following gauge parameter version
of~\eqref{YC},
\begin{equation}
  (\os_i\dl{\os_j}+b_i\dl{b_j})\chi^{(k)}=\, s_i
  \,\delta_{ij}\,\chi^{(k)} 
\quad (i\geq j)\,, \qquad \qquad \os_i \cdot \dl{x} \chi^{(k)}=0. 
\end{equation}

\subsubsection{Obstructions at the level of the action}

The BRST state cohomology $H^{g}(\Qbrst)$ for these systems has been
computed in \cite{Alkalaev:2008gi} and shown to be isomorphic to
$H^{g}(\Qred)$ through the elimination of contractible pairs. It follows
that $H^{g}(\Qred)$ is a module of the global symmetry algebra. It is
particularly convenient to consider $H^{-p}(\Qbrst)$. Recall that $p$ is
the number of nonvanishing spin labels $s_i$, and hence is the maximal
homogeneity degree in $b_i$, i.e., the number of rows in the YT
describing the field. Indeed, as there are no nonzero elements in
degree $<-p$, the coboundary condition is trivial and $H^{-p}(\Qbrst)$ is
given by $\chi^{(p)}=b_1\ldots b_p\, \xi(x,p)$ where $\xi$ satisfies
\begin{equation}
  \os_i\cdot \dl{\os_j}\xi=\delta_{ij}\,(s_i-1)\,\xi  
\quad (i \geq j) \,, \qquad \os_i \cdot \dl{x} \xi=0\,.
\end{equation} 
along with~\eqref{TT}.

$H^{-p}(\Qbrst)$ is a Poincar\'e-module composed of irreducible
Lorentz-modules associated with YT $(s_1-1,\ldots, s_p-1, k)$ where
$0\leq k \leq s_p-1$ and $s_1\geq \ldots \geq
s_p$~\cite{Alkalaev:2008gi}. These modules can also be inferred from
the unfolded formulation~\cite{Skvortsov:2008sh}. The subspace $V_0
\subset H^{-p}(\Qbrst)$ annihilated by Poincar\'e translations is an
irreducible module with weights $s_1-1,\ldots, s_p-1$.

Repeating the arguments based on the generalized Verma module induced
from this $o(d-1,1)$-module one finds that $-\Delta_L\geq s_1-1$ and
the decomposition of the corresponding finite-dimensional
$o(d,2)$-module $-\Delta_L,s_1-1,\ldots,s_p-1$ necessarily contains
modules not present in the starting point Poincar\'e-module except if
$s_1=\ldots=s_p=1$ and $\Delta_L=0$. The gauge field with such a
$H^{-p}(\Qbrst)$ is a totally-antisymmetric field of rank $p$.

Again, this information infered just from $H^{-p}(\Qbrst)$ is not
enough to conclude for which $p$ a totally antisymmetric field is
conformal in $d$-dimensional Minkowski spacetime. Similar to the case
of totally symmetric fields, if the system is Lagrangian, the
Lagrangian is of second order in derivatives, so that one gets $d/2-1$
as the weight for the gauge field itself. Furthermore,
$H^{-p}(\Qbrst)$ corresponds to $p$-th level reducibility identities
with each level involving first order operators, which gives
$\Delta_L=\frac{d}{2}-1-p$ for the conformal weight of $V_0$. Together
with $\Delta_L=0$ obtained above, this shows that the only remaining
candidates are antisymmetric fields of rank $p=\frac{d}{2}-1$ in
(even) dimension $d$, which are indeed known to be conformal.

\subsubsection{Obstructions at the level of equations of motion}

If one is only interested in EOM symmetries of gauge fields then, in
order to see that only rank $\frac{d}{2}-1$ totally antisymmetric
gauge fields are conformal, one needs to consider $H^0(\brst)$ as
well, i.e., the space of gauge-inequivalent solutions to the EOM. For
such fields, this space is a conformal module for $p=\frac{d}{2}-1$
~\cite{Siegel:1988gd,Angelopoulos:1997ij,Shaynkman:2004vu}.

As we discussed in~\bref{sec:constr-curv-form}, the analysis of
$H^0(\brst)$ is equivalent to an analysis in terms of curvatures
because $H^0(\brst)$ is the same for the gauge field and its
formulation in terms of curvatures. Let us then briefly review the
known results concerning fields in Minkowski spacetime that are
conformal in terms of curvatures or, more precisely, which Poincar\'e
irreducible non-gauge fields in Minkowski spacetime are conformal.

It turns out that in odd $d$ only a massless scalar and spinor field
are conformal, while in even $d$ there are in addition ``spinning''
singletons. The latter are fields described by irreducible Lorentz
tensors associated to rectangular YT of height $\frac{d}{2}$, which
are in particular, traceless and (anti)-selfdual. In fact, they
correspond to the massless gauge fields with $p=\frac{d-2}{2}$ and
$s_1=\ldots=s_p=s$, when formulated in terms of curvatures. More
precisely, the above irreducible tensors are the gauge-invariant
generalized Weyl tensors of these gauge fields. Their conformal
invariance was originally shown by identifying those Poincar\'e irreps
that lift to conformal
ones~\cite{Siegel:1988gd,Angelopoulos:1997ij}. In terms of EOM
symmetries this follows from the results of~\cite{Shaynkman:2004vu},
while a manifestly local and conformal formulation of these bosonic
spinning singletons in terms of curvatures was constructed
in~\cite{Bekaert:2009fg}.

This completes our discussion of possible conformal invariance of
bosonic gauge fields on Minkowski spacetime. The extension to
fermionic fields is straightforward using
e.g.~\cite{Campoleoni:2009gs,Skvortsov:2010nh}. Note that we have not
explicitly discussed massive nor continuous spin representations as
they cannot be conformal. This follows essentially from the fact that
both of them involve a dimensionful parameter.

\section{Gauge fields in anti-de Sitter spacetime}

\label{sec:AdSfields}

\subsection{Maximal-depth partially-massless fields in 4d}
\label{amb-PM-FT}
We begin the analysis of possible conformal invariance of $AdS$ gauge
fields with the relatively simple, but not so well-known example of
totally symmetric partially massless (PM)
fields~\cite{Deser:1983tm,Deser:1983mm,Deser:2001us,Deser:2001pe,Skvortsov:2006at}
of maximal depth $t=s$.  In this case the gauge parameter is a
scalar. In terms of the $d+1$-dimensional ambient space with
coordinates $X^B$, $(B=0,1,\cdots,d-1,d)$ and flat metric
$\eta_{AB}=\mbox{diag}(-,+,\cdots,+,-)$, anti-de Sitter spacetime
$AdS_d$ is the hyperboloid $X\cdot X+1=0$. In these terms, the gauge
field is encoded in the generating function $\phi(X,\Os)$ subject
to~\cite{Grigoriev:2011gp,Alkalaev:2011zv,Joung:2012rv,Bekaert:2013zya}
\begin{equation}
\begin{gathered}
  (X\cdot\dl{X}+1)\Phi=0\,, \qquad X\cdot \dl{\Os}\Phi=0\,, \qquad
  (\Os\cdot \dl{\Os}-s)\Phi=0\,,\\  
 \dl{X}\cdot \dl{X}\Phi=\dl{X}\cdot \dl{\Os}\Phi=\dl{\Os}\cdot \dl{\Os}\Phi=0\,,
 \end{gathered}
\end{equation} 
and the gauge transformations
\begin{equation}\label{PMs=tgtransfo}
 \delta_\chi \Phi=(\Os\cdot \dl{X})^s\chi\,, \quad (X\cdot\dl{X}-s+1)\chi=0\,,
 %\quad (\Os\cdot\dl{X})^s\Phi=0\,, 
 \quad \dl{X}\cdot \dl{X}\chi=0\,.
\end{equation} 
The variables $\Os^B$, $B=0,\ldots,d$ are introduced to contract
tensor indices. Note that $\chi$ is $\Os$-independent.

Just like in the case of Minkowski spacetime fields considered above,
it is convenient to introduce a ghost variable $b$ and consider the
space of states  of the form $\Phi(X,A)+b\,\chi(X,A)$ with BRST
operator $\Qred=(\Os\cdot \dl{X})^s\chi\dl{b}$ implementing the above
gauge equivalence. Although the space of gauge parameters is subject
to differential constraints, such a formulation is equivalent to a
formulation based on a suitable BRST operator $\Qbrst$ with free gauge
parameters~\cite{Alkalaev:2009vm}. In particular, $H(\Qbrst)\simeq
H(\Qred)$.

The global reducibility parameters $H^{-1}(\brst)$ are determined by
$(\Os\cdot \dl{X})^s{\chi_0}=0$.  This condition requires $\chi$ to be
polynomial in $X$.  The first condition in \eqref{PMs=tgtransfo} fixes
the homogeneity of the polynomial to be $s-1$. Finally, the second
condition allows one to conclude that $H^{-1}(\brst)$ is the space of
totally traceless rank $s-1$ tensors in $d+1$ dimensions. This is an
irreducible module of the $AdS_{d}$ isometry algebra $o(d-1,2)$. Note
that irreducibility implies that there can be no gauge symmetries for
the gauge parameters in this system.

Following the same idea as before, let us try to check if this
$o(d-1,2)$-module can also be an $o(d,2)$-module. Leaving the rigorous
and general proof for the next section, let us present a simple
heuristic proof. Observe that all finite-dimensional $o(d,2)$-modules
described by 1-row Young tableaux are simply exhausted by totally
traceless fixed rank totally symmetric tensors in $d+2$ dimensions,
rather than in $d+1$ dimensions as above.  One then concludes that the
two spaces do not coincide unless $s=1$.  In particular this implies
that depth $t=s$ PM fields in $4d$ are not conformal as gauge systems,
i.e. in terms of potentials, unless $s=1$, in which case it is the
usual Maxwell field.

Although maximal-depth PM fields in 4 dimensions are not conformal in
general, there exist very similar maximal-depth conformal gauge
fields.  For $s=1$ they coincide with the Maxwell field, for $s=2$
they were originally found in~\cite{Deser:1983mm}, and for generic $s$
in~\cite{Erdmenger:1997wy}. They can be seen as higher-depth
generalization of usual conformal gauge fields~\cite{Fradkin:1985am},
and hence, we call them maximal-depth FT fields below. They belong to
the class of conformal gauge fields considered
in~\cite{Vasiliev:2009ck}. Recently they were identified
with boundary values of the $AdS_5$ maximal-depth PM fields~\cite{Bekaert:2013zya}.
In $d=4$ these fields have second order equations of motion and gauge
transformation of order $s$ in the derivatives. More precisely, the
flat spacetime Lagrangian for traceless $\varphi_{\mu_1\ldots \mu_s}(x)$
reads as~\cite{Erdmenger:1997wy}
\begin{equation}
 L=\d^\nu\varphi^{\mu_1\ldots \mu_s}\d_\nu\varphi_{\mu_1\ldots \mu_s}-\frac{2s}{s+1}
 \d_\nu\varphi^{\nu\mu_2\ldots \mu_s}\d^\lambda\varphi_{\lambda\mu_2\ldots \mu_s}
\end{equation} 
and is invariant under $\delta \varphi_{\mu_1\ldots
  \mu_s}=\d_{\mu_1}\ldots \d_{\mu_s}\chi-\text{traces}$. Thanks to
conformal invariance, they can be seen as fields on any conformally
flat space and, in particular, on $AdS_{4}$. A natural question is
then what their relationship to the maximal-depth PM fields on the
same spacetime precisely is.

To answer this question, let us consider again global
reducibilities. Using the ambient formulation
of~\cite{Bekaert:2013zya}, the space of reducibilities can be described
in terms of polynomials in $d+2$-variables $X^M$ satisfying
\begin{equation}
 X^M \dl{X^M} \Phi=(s-1)\Phi\,, \qquad 
 \eta^{MN}\dl{X^M}\dl{X^N} \Phi=0 \,.
\end{equation} 
This subspace is determined by the same equations as $H^{-1}(\brst)$
above but in $d+2$ dimensions. Unless $s=1$ these spaces do not
coincide.  In Section~\bref{sec:spin2} we explicitly compare these two
fields in the first nontrivial case of $s=2$.

As far as totally symmetric PM fields of maximal depth are concerned,
one can wonder if, similarly to Fronsdal fields in 4 dimensions, the
equations of motion are conformal in terms of curvatures. To answer
this question we use the formulation in terms of curvatures proposed
in~\cite{Deser:2006zx} (see e.g. Sec.~\bref{sec:spin2} for the
simplest non trivial example of $s=2$). If these systems were
conformal, one could equally well rewrite them in flat Minkowski
spacetime using a Weyl transformation. As the flat limit for these AdS
systems is regular, its Weyl transformation to flat space should
coincide with its naive flat limit obtained by putting the
cosmological constant to zero. More precisely, for the flat limit of a
PM maximal-depth field, the fundamental field is an irreducible
Lorentz tensor $F_{\mu_1\ldots\mu_s|\nu}$, i.e., it is symmetric over
all $\mu$ indices and such that the complete symmetrization over all
lower indices gives zero. It then follows from the classification
results of~\cite{Shaynkman:2004vu} that, for such a Lorentz tensor
field labelled by a ``hook'' YT $(s,1)$\,, there are only two
conformal equations which are first-order in derivatives and a
rank-$s$ Lorentz tensor: one is a totally symmetric rank-$s$ Lorentz
tensor with conformal weight $2$ while the other one is labelled by a
hook YT $(s-1,1)$ and has conformal weight $s+3$. The former equation
corresponds to the curvature formulation of a maximal-depth conformal
gauge field which differs from the corresponding PM field unless
$s=1$. This difference is explicitly illustrated on the example of
$s=2$ in Sec.~\bref{sec:spin2} below. The latter equation also differs
from the corresponding PM field since in particular, the curvature has
a different conformal weight. In conclusion:~\footnote{At first
  glance, this conclusion differs from \cite{Deser:2004ji} but this
  paper is based on different assumptions and makes use of a different
  definition of symmetries. In particular, the $o(4,2)$ symmetry
  discussed in \cite{Deser:2004ji} does not seem to correspond to
  standard conformal spacetime transformations.}

{\em Maximal-depth PM field with $s>1$ are not conformal, neither in
  terms of potentials, nor in terms of curvatures.}

\subsection{Generic partially-massless gauge fields in AdS}

A partially massless bosonic gauge field in $AdS_d$ is determined by a
finite-dimensional module of $o(d-1)$ with weights (spins)
$s_1,\ldots,s_r$. Here $r=[\frac{d-1}{2}]$ is the rank of a rotation
subalgebra $o(d-1)$, while $s_1\geq\ldots \geq s_r$, $p$ and $t$ are integer
parameters, $1\leq p \leq r$ and $1 \leq t \leq s_p-s_{p+1}$\,.
%, with $s_{p+1}=0$ for $p>r$. 
This corresponds to a (partially)-massless
field of spin $s_1,\ldots,s_r$ with depth-$t$ gauge transformation
associated to the $p$-th row.  More details can be found 
in~\cite{Metsaev:1995re,Alkalaev:2003qv,Skvortsov:2009zu,%
  Alkalaev:2009vm,Alkalaev:2011zv}.

The BRST first-quantized description for a generic bosonic gauge field
on AdS has been constructed in~\cite{Alkalaev:2009vm,Alkalaev:2011zv}
(see
also~\cite{Barnich:2006pc,Alkalaev:2003qv,Skvortsov:2009zu,Skvortsov:2009nv,%
  Boulanger:2008up} for earlier related work).  The nontrivial
$H^{-i}(\brst)$ are in degree $0$ and $p$.  For an irreducible
(partially)-massless field, the space $H^{-p}(\brst)$ is a
finite-dimensional irreducible $o(d-1,2)$-module with highest weight
$s_1-1,\ldots,s_{p-1}-1, s_p-1, s_{p}-t, s_{p+1},\ldots,s_{r}$,
i.e. the module described by the Young diagram with the lengths of
rows given by:~\footnote{It is this module where a $p$-form field
  takes values in the unfolded
  description~\cite{Alkalaev:2003qv,Skvortsov:2009zu} of AdS gauge
  fields.}
\begin{equation}
\label{hp}
s_1 -1\geq \ldots \geq s_{p-1}-1
\geq s_p-1 \geq s_{p}-t
\geq s_{p+1}\geq \ldots \geq s_{r}\;.
\end{equation}
Note the row of length $s_p-1$ in the middle of the diagram and a
subsequent row of length $s_p-t$. For instance, for $d=4$ and $t=1$,
one gets the familiar 2-row rectangular tableaux of length
$s_1-1$. Note that $r=1$ in this case.

According to the $o(d+2)\downarrow o(d+1)$ branching rules summarized
in the next subsection, if module~\eqref{hp} is nontrivial, it can be
lifted to $o(d,2)$ iff $d$ is even and this Young tableau is
rectangular of height $\frac{d}{2}$. This condition resricts $s_i$ in such a way
that $s_1-1=s_2-1\ldots=s_p-1=s_{p+1}=\ldots=s_r$ so that according to~\cite{Metsaev:1995jp}
the field belongs to the class of unitary mixed-symmetry fields. In particular, 
$t=1$ so that mixed symmetry PM fields cannot be conformal in general.

To obtain further restrictions one has to consider $H^0(\brst)$ as well. 
According to the analysis of~\cite{Metsaev:1995jp} unitary AdS fields may admit 
conformal symmetry only for  $s_1=\ldots=s_r$ (in particular $p=r$) and $d$ even. If 
we restrict ourselves to the case $p=r$, $H^{-p}(\brst)$ is associated to a 
spinning singleton \cite{Bracken:1982ny,Siegel:1988gd,Metsaev:1995jp}. As an 
$o(d,2)$ module, $H^{-p}(\brst)$ is a finite-dimensional module described by a 
rectangular tableau of height $\frac{d}{2}+1$ and length $s-1$. The module is 
realized by (anti)-selfdual tensors of this symmetry type in $d+2$-dimensions. 
In dimensions $d$ different than $2~ mod ~4$ however, modules of this sort are 
necessarily complex as the (anti-) selfduality condition does not have real 
solutions in such dimensions\footnote{This is in agreement 
with~\cite{Vasiliev:2001zy} where the conformal invariance of doubled 
(complexified) sets of totally-symmetric fields in $AdS_4$ was put forward.}. In 
particular, this implies that in $AdS_d$ with $d=4,8,\ldots$ real fields can be 
conformal in terms of potentials only for $s_1=\ldots =s_r=1$, i.e., when the 
module is trivial. These are totally antisymmetric fields of maximal rank which 
are known to be conformal for even $d$.

An interesting question is whether spinning singletons in $AdS_d$ with
$d\geq 6$, $d=2 ~mod~4$ and $s>1$ can be conformal in terms of
potentials. Note that those with $s=1$ are conformal in terms of
potentials, while they all are known to be conformal in terms of
curvatures. The necessary condition advocated here does not exclude
this possibility and resolving the issue requires further study.

\subsection{Branching rules for modules of the orthogonal algebras}
\label{sec:branching-rules}

The branching rules of a Lie algebra $\mathfrak{g}$ describe the
decomposition of its irreps restricted to a subalgebra
$\mathfrak{h}$. We will be interested in the very exceptional case
when the $\mathfrak{g}$-irrep remains irreducible under the
restriction $\mathfrak{g}\downarrow\mathfrak{h}$, i.e., when the
decomposition contains only a single $\mathfrak{h}$-irrep with
multiplicity one. The trivial representation is an obvious example of
such an irrep. The branching rules of classical algebras are
well-known for finite dimensional irreps while the problem is
obviously more involved for infinite-dimensional ones.

The importance of branching rules for our purpose is the following
fact: \textit{An $\mathfrak{h}$-irrep can be lifted to a
  $\mathfrak{g}$-irrep if and only if this $\mathfrak{h}$-irrep is the
  only irrep appearing in the restriction
  $\mathfrak{g}\downarrow\mathfrak{h}$ of the %lifted
  $\mathfrak{g}$-irrep.} In other words, there is a one-to-one
correspondence between the $\mathfrak{h}$-irreps that can be lifted to
$\mathfrak{g}$-irreps and the $\mathfrak{g}$-irreps that remains
irreducible under the restriction
$\mathfrak{g}\downarrow\mathfrak{h}$.

To see which finite-dimensional $o(d-1,2)$-modules can be lifted to
$o(d,2)$, we recall the basic facts on $o(d)\downarrow o(d-1)$
branching rules. The finite-dimensional irreducible ${o}(d)$-module
characterized by the dominant integral ${o}(d)$-weight $\vec
s\equiv(s_1,\ldots,%s_{r-1},
s_r)$ will be denoted by $\cD_{{o}(d)}(\vec s)$. Here $r$ denotes the
rank of $o(d)$, i.e., the integer part of $d/2$. The ``spin'' labels
of the weight $r$-vector $\vec s$ are either all integers or all
half-integers, and they satisfy
\begin{eqnarray}
&&s_1 \geqslant \ldots \geqslant s_r \geqslant  0\quad \mbox{for}\quad
d = 2r +1\;, 
\nonumber \\
&&s_1 \geqslant \ldots \geqslant s_{r-1} \geqslant \vert s_r \vert
\quad \mbox{for}\quad d = 2r\;. 
\end{eqnarray}
When $d=2r$, the last label $s_{r}$ can be positive or negative. The
integer part of the (absolute values) of the components in $\vec s$
define a Young diagram where each spin label gives the length of the
corresponding row.

The classical branching rules for the restriction ${o}(d)\downarrow
o(d-1)$ of finite-dimensional irreducible modules can be expressed as
follows:
\begin{equation}
\cD_{{o}(d)}(\vec s)\,\downarrow\,\bigoplus\limits_{\vec t%\vec s
}\cD_{{o}(d-1)}(\vec t)\,,
\label{branchod}
\end{equation}
where the direct sum is over all ${o}(d-1)$-weights $\vec t$ such that
\begin{eqnarray}
&&s_1 \geqslant t_1 \geqslant\ldots\geqslant s_{r-1} \geqslant t_{r-1}
\geqslant s_r \geqslant |t_r| \quad \mbox{for}\quad d = 2r +1\;,
\label{br1} \\
&&s_1 \geqslant t_1 \geqslant\ldots\geqslant s_{r-1} \geqslant t_{r-1}
\geqslant |s_r| \quad \mbox{for}\quad d = 2r\;,\label{br2}
\end{eqnarray}
with entries in $\vec s$ and $\vec t$ which are simultaneously all
integers or all half-integers.

\begin{lemma}\label{nonbranchingo(d)}
  A nontrivial irreducible ${o}(d)$-module $\cD_{{o}(d)}(\vec s)$
  remains irreducible after its restriction to ${o}(d-1)$ if and only
  if $d=2r$ and $s_1=\ldots=s_{r-1}=|s_r|$, i.e., if it is described by
  a rectangular Young diagram of height $d/2$\,.
\end{lemma}

\proof{The branching rules \eqref{br1} and \eqref{br2} imply the
  following chain of inequalities $s_1 \geqslant t_1
  \geqslant\ldots\geqslant s_{r-1} \geqslant t_{r-1} \geqslant |s_r|$
  which are valid in any $d$. One can see that a necessary condition
  in order to have a single allowed set of components $t_1$, ....,
  $t_{r-1}$ is that $s_1=\ldots=s_{r-1}=|s_r|$. For $d=2r$, this fixes
  uniquely $\vec t$ to be the ($r-1$)-vector (since ${o}(d-1)$ has
  rank $r-1$) such that $t_1=\ldots=t_{r-1}=|s_r|$.  For $d=2r+1$,
  inspecting the last inequality $s_r \geqslant |t_r|$ in the
  branching rule \eqref{br1}, one can see that $s_r$ must vanish in
  order to have a single allowed component $t_r$. This implies that
  the trivial irreducible ${o}(d)$-module $\cD_{{o}(d)}(\vec 0)$ is
  the only one that remains irreducible after restriction to
  ${o}(d-1)$ for $d$ odd.}

An obvious corollary is that, if one performs two such branchings, the
only irreducible ${o}(d)$-module which remains irreducible after its
restriction to ${o}(d-2)$ is the trivial module.

\subsection{Explicit spin 2 examples}
\label{sec:spin2}

To illustrate the difference between $s=t=2$ PM field and FT field in
4d, let us work in terms of tangent tensors.

\subsubsection{Maximal-depth partially-massless spin-2 field in 4d}

\paragraph{$s=t=2$ PM field in $d=4$ in terms of potentials:}

Following~\cite{Deser:1983tm,Deser:1983mm}, the equations of motion for a $s=t=2$ PM
field in 4d are
\begin{multline}
\label{eq:st2PM}
 (\nabla^2+4\mu^2)\varphi_{\mu\nu}-(\nabla_\mu
 \nabla^\rho\varphi_{\rho\nu}+\nabla_\nu
 \nabla^\rho\varphi_{\rho\mu})+\nabla_\mu\nabla_\nu \varphi^\prime 
 -
 \\
 -g_{\mu\nu}\left((\nabla^2+\mu^2)\varphi^\prime-\nabla^\rho\nabla^\sigma
   \varphi_{\rho\sigma}\right)=0\,, 
\end{multline} 
where $\varphi^\prime \equiv g^{\mu\nu}\varphi_{\mu\nu}$. Here $\mu=L^{-1}$
the inverse AdS radius so that e.g. $R_{\mu\nu\rho\sigma}=-\mu^2
(g_{\mu\rho}g_{\nu\sigma}-g_{\nu\rho}g_{\mu\sigma})$. The equations
are invariant under the following gauge
symmetry
\begin{equation}
\label{PM-gs}
 \delta_\xi \varphi_{\mu\nu}=(\nabla_\mu\nabla_\nu-\mu^2 g_{\mu\nu})\xi
\end{equation} 
with unconstrained scalar parameter $\xi(x)$. 

Equations~\eqref{eq:st2PM} have differential consequences of first
order~\cite{Deser:2001us}. Applying $\nabla^\mu$ to both sides
of~\eqref{eq:st2PM} one finds
\begin{equation}
 \nabla^\mu\varphi_{\mu\nu}-\nabla_\nu \varphi^\prime=0\,,\qquad
 \varphi^\prime\equiv g^{\mu\nu}\varphi_{\mu\nu}\,. 
\end{equation} 

Let us also present the partially gauge fixed version of this
system. Namely, let us consider the gauge condition
$\varphi^\prime=0$. Its variation under a gauge transformation is given
by
\begin{equation}
 \delta \varphi^\prime=(\nabla^2-4\mu^2)\xi\,,
\end{equation} 
so that the gauge is reachable. Indeed, in the context of jet-spaces,
any element is in the image of $\nabla^2$. The gauge fixed system
reads
\begin{equation}
\label{PM-gf}
 \begin{gathered}
(\nabla^2+4\mu^2) \varphi_{\mu\nu}=0\,,\quad \nabla^\mu
\varphi_{\mu\nu}=0\,,\quad g^{\mu\nu}\varphi_{\mu\nu}=0\,,  \\ 
\delta\varphi_{\mu\nu}=(\nabla_\mu\nabla_\nu-\mu^2 g_{\mu\nu})\xi\,,
\qquad (\nabla^2-4\mu^2)\xi=0 \,.
\end{gathered}
\end{equation} 
This formulation can be rewritten in ambient terms by identifying
$\varphi_{\mu\nu}$ with the pullback of ambient $\varphi_{AB}$ satisfying
$X^A\varphi_{AB}=0$, $(X\cdot \dl{X}+1)\varphi_{AB}=0$, and similarly for
the gauge parameter.

The space of global reducibilities is determined by
$\delta_\xi\varphi_{\mu\nu}=0$. The consequence $g^{\mu\nu}\delta_\xi
\varphi_{\mu\nu}=0$ reads explicitly
\begin{equation}
(\nabla^2-4\mu^2)\xi=0\,.
\end{equation} 
Let us identify $\xi$ as the pullback of $\Xi(X)$ defined on ambient
space $\fR^{3+2}$ and satisfying $(X\cdot \dl{X}-1)\Xi=0$,
$\dl{X}\cdot \dl{X} \Xi=0$. In terms of $\Xi$, the gauge
transformation is $\d_A\d_B \Xi$ and hence $\Xi$ must be polynomial.
One concludes that $\Xi=\xi_A X^A$, so that reducibilities are
parametrized by $d+1$ dimensional ambient vectors.

\paragraph{$s=t=2$ PM field in $d=4$ in terms of curvatures:}

Following~\cite{Deser:2006zx}, the curvature is given by
\begin{equation}
 F_{\mu\nu|\rho}=\nabla_\mu \varphi_{\nu\rho}-\nabla_\nu \varphi_{\mu\rho}\,.
\end{equation} 
In terms of $F_{\mu\nu|\rho}$, equations of motion~\eqref{eq:st2PM}
take the form
\begin{equation}
\label{PM-eom-curv}
 \nabla^\rho F_{\rho (\mu|\nu)}-g_{\mu\nu} \nabla^\rho F^\prime_\rho +
 \nabla_{(\mu} F^\prime_{\nu )}=0\,, 
\end{equation} 
where $F^\prime_\mu=F_{\mu\rho |\nu} g^{\rho\nu}$ and $X_{(a}
Y_{b)}=\half (X_a Y_b+X_b Y_a)$. In this form, the equations of motion
follow from the Lagrangian~\cite{Deser:2006zx}:
\begin{equation}
 L^{PM}=F_{\mu\nu|\rho} F^{\mu\nu|\rho }+F^{\prime\,\nu }F^\prime_{\nu }\,.
\end{equation} 

If one treats $F_{\mu\nu|\rho}$ as the fundamental field, one also
needs to add algebraic conditions and Bianchi identities so that the
complete set of equations becomes
\begin{gather}
 \label{PM-alg}
 F_{\mu\nu|\rho}=-F_{\nu\mu|\rho}\,, \qquad 
  F_{[\mu\nu|\rho]}=0\,, \qquad F_{\mu\nu|\rho}g^{\nu\rho}=0\,,\\
 \label{PM-eq}
  \nabla^\mu F_{\mu\nu|\rho}=0\,, \qquad \nabla_{[\sigma }F_{\mu\nu ] |\rho}=0\,.
\end{gather}
Note that if $F_{\nu\mu|\rho}$ is (anti)-selfdual the last two
equations are equivalent.

\subsubsection{Maximal-depth Fradkin-Tseytlin spin-2 field in 4d}

\paragraph{$s=t=2$ FT field in $d=4$ in terms of potentials:}

Another related system in 4d was also proposed
in~\cite{Deser:1983tm,Deser:1983mm} (see also references therein). The
equations of motion have the form
\begin{equation}
 (\nabla^2+4\mu^2)\varphi_{\mu\nu}-\frac{2}{3}(\nabla_\mu
 \nabla^\rho\varphi_{\rho\nu}+\nabla_\nu
 \nabla^\rho\varphi_{\rho\mu})+\frac{1}{3}g_{\mu\nu} \nabla^\rho\nabla^\sigma
 \varphi_{\rho\sigma}=0 
\end{equation} 
and $g^{\mu\nu}\varphi_{\mu\nu}=0$. The gauge law is 
\begin{equation}
 \delta_\xi \varphi_{\mu\nu}=(\nabla_\mu\nabla_\nu-\frac{1}{4} g_{\mu\nu}
 \nabla^2)\xi 
\end{equation} 
with $\xi$ unconstrained. This system is conformal and can be
identified \cite{Bekaert:2013zya} with the boundary value of the
$t=s=2$ PM field on $AdS_5$ .

In contrast to the $s=t=2$ PM field considered above, the gauge
$\nabla^\mu \varphi_{\mu\nu}=0$ is not reachable in general. On the
contrary, $V_\mu:=\nabla^\mu \varphi_{\mu\nu}$ satisfy Maxwell's equations
and transform as $\delta
V_\mu=\frac{3}{4}\nabla_\mu(\nabla^2-4\mu^2)\xi$.

To see what this system describes, let us decompose $\varphi_{\mu\nu}$
(in a nonlocal way) into $\varphi^0_{\mu\nu}$ satisfying $\nabla^\mu
\varphi^0_{\mu\nu}=0$ and $V_\mu$ describing the rest.  The equations for
$\varphi_0$ reduce to~\eqref{PM-gf}, so that a FT field with $s=t=2$
decomposes into a PM field $\varphi^0$ with $s=t=2$ and a Maxwell field $V$ with
$s=t=1$.

The space of global reducibilities is given by solutions to
$(\nabla_\mu\nabla_\nu-\frac{1}{4} g_{\mu\nu} \nabla^2)\xi=0$. Let us
consider first the consequence $\nabla^\mu \delta_\xi(
\varphi_{\mu\nu})=0$, or explicitly,
\begin{equation}
 \nabla^\mu \delta_\xi(
 \varphi_{\mu\nu})=\frac{3}{4}\nabla_\nu(\nabla^2-4\mu^2)\xi=0\,. 
\end{equation} 
The general solution to this equation has the form $\xi=a+\xi_0$ where
$a$ is constant and $\xi_0$ is a general solution to
$(\nabla^2-4\mu^2)\xi_0=0$. In turn, just like in the case of a PM
field, it is convenient to represent $\xi$ as the pullback to the
hyperboloid of $\Xi_0$ defined on $\fR^{3+2}$ and satisfying
$\dl{X}\cdot \dl{X} \Xi_0=0$, $(X\cdot \dl{X}-1)\Xi_0=0$. In terms of
the ambient space, conditions $\delta_{\xi_0}\varphi_{\mu\nu}=0$ take the
form $\d_A \d_B \Xi_0=0$ where $(\nabla^2-4\mu^2)\xi_0=0$ has been
taken into account. So the solution is again given by $\Xi_0=\xi_A
X^A$. Putting everything together, the general solution for $\xi$ is
$\xi=a+\xi_A X^A(x)$ and the space of reducibilities is 6-dimensional,
confirming the conclusion of the manifestly conformal
considerations of Section~\bref{amb-PM-FT}. Let us stress that in
contrast to Section~\bref{amb-PM-FT},  we now have not assumed that
conformal symmetry is realized on gauge parameters.

\paragraph{$s=t=2$ FT field in $d=4$ in terms of curvatures:}

The traceless component of the curvature is
\begin{equation}
 \tilde F_{\mu\nu|\rho}=\nabla_\mu \varphi_{\nu\rho}-\nabla_\nu
 \varphi_{\mu\rho}-\frac{1}{3}g_{\mu\rho}\nabla^\alpha \varphi_{\alpha
   \nu}+\frac{1}{3}g_{\nu\rho}\nabla^\alpha \varphi_{\alpha \mu}\,. 
\end{equation} 
In terms of $\tilde F$, the equation of motion take the form
\begin{equation}
\label{eq-FT}
 \nabla^\mu \tilde F_{\mu(\nu|\rho)}=0\,.
\end{equation} 
They follow from the Lagrangian
\begin{equation}
 L^{FT}=\half \tilde F_{\mu\nu|\rho}\tilde F^{\mu\nu|\rho}\,.
\end{equation} 

If one treats $\tilde F_{\mu\nu|\rho}$ as the fundamental fields, the
complete set of equations is
\begin{gather}
\label{FT-alg}  {\tilde F}_{\mu\nu|\rho}=-{\tilde F}_{\nu\mu|\rho}\,, \qquad 
 {\tilde F}_{[\mu\nu|\rho]}=0\,, \qquad {\tilde F}_{\mu\nu|\rho}g^{\nu\rho}=0\,,\\
 \label{FT-eq} \nabla^\mu {\tilde F}_{\mu(\nu|\rho)}=0\,, \qquad
 \nabla_{[\sigma }{\tilde F}_{\mu\nu ]
   |\rho}=g_{\rho[\sigma}A_{\mu\nu]}\,, 
\end{gather}
where $A_{\mu\nu}$ is an antisymmetric tensor. The last equations can
be written as $\cP( \nabla_{[\sigma}\tilde F_{\mu\nu]|\rho})=0$, where
$\cP$ denotes the projector to the totally traceless component. Note
that if $\tilde F_{\nu\mu|\rho}$ is (anti)-selfdual, the last two
equations are equivalent. 

By comparing~\eqref{FT-alg}, \eqref{FT-eq} to \eqref{PM-alg},
\eqref{PM-eq}, one observes that the $s=t=2$ FT equations of motion
are a subset of the $s=t=2$ PM equations. Therefore, the space of
solutions of the $s=t=2$ PM equations is a subspace of the $s=t=2$ FT
one. Indeed, the former is an $o(d-1,2)$-submodule of the latter. The
crucial point is that, nevertheless, the former is \textit{not} an
$o(d,2)$-submodule of the latter because the extra equations of the
$s=t=2$ PM field are \textit{not} conformally invariant for the
conformal weight of the $s=t=2$ FT field. The same remains true for
$s>2$.

\section{Conclusion}
\label{sec:conclusion}

In this work we have studied structural properties of global
symmetries in gauge systems. In particular, in the context of the
BV-BRST approach, we have shown that BRST cohomology in the space of
local functionals, $H^{-p-1}(s)$, as well as BRST-state cohomology
$H^{-p}(\brst)$ in the case of linear systems, are necessarily modules
over any subalgebra of the algebra of global symmetries.

Of special importance are ``global reducibility parameters'' which
correspond to these cohomology groups for $p\geq 1$. In contrast to
BRST cohomology groups in other ghost numbers, global reducibilities
are typically finite-dimensional. This makes them especially useful in
order to constrain global symmetries since the analysis then only
requires standard tools from representation theory. Surprisingly, in
the particular examples where we study which (A)dS or Poincar\'e gauge
fields admit conformal symmetry, this analysis is powerful enough to
rule out most of the candidates, without analyzing the space of
solutions.

Our approach is closely related to the unfolded formalism.  Namely, in
the unfolded approach, the construction of gauge field begins with the
choice of a finite-dimensional module and with differential forms
taking values in this module. The detailed relationship can be
established using a parent approach which allows one to systematically
construct an unfolded formulation starting from the BV-BRST
formulation, respectively its BRST first quantized formulation for
linear theories: the space where the $p$-form fields take values in
the minimal unfolded formulation can then be shown to coincide with
$H^{-p}(\brst)$, and hence with order $p$ global reducibility
parameters.

\section*{Acknowledgements}

%\label{sec:acknowledgements}

\addcontentsline{toc}{section}{Acknowledgments}

The work of GB is supported by the Fund for Scientific Research-FNRS
(Belgium), by IISN-Belgium, and by ``Communaut\'e fran\c caise de
Belgique - Actions de Recherche Concert\'ees''.  The research of XB
and MG was supported by the Russian Science Foundation grant
14-42-00047 in association with Lebedev Physical Institute.  XB is
grateful to Sogang University for hospitality while this work was
being completed. XB and MG would like to thank K.B.~Alkalaev,
E.~Joung, R.~Metsaev, Y.~Nakayama, I.~Tipunin, A.~Tseytlin,
M.A.~Vasiliev, A.~Verbovetsky and especially O.~Shaynkman for valuable
discussions.

\vspace{2mm}
\noindent\textbf{Note added:} While the present paper was in
preparation, reference \cite{Beccaria:2015vaa} appeared on
\texttt{arXiv}, which also discusses $4$-dimensional maximal-depth
Fradkin-Tseytlin fields.  
\vspace{2mm}

{\small

\addtolength{\baselineskip}{-4.2pt}
\addtolength{\parskip}{-4pt}

%\fi
%\bibliography{master}
%\bibliography{/Users/gbarnich/Dropbox/Literature/master}
%\bibliography{C:/Users/Glenn/Dropbox/Literature/master}
%\bibliography{/home/maxim/Documents/HSmaster}

\providecommand{\href}[2]{#2}\begingroup\raggedright\endgroup

}
\end{document}